\pdfoutput=1

\documentclass[11pt,twoside,a4paper,cmspaper,final,collab]{cms-tdr}
\def\svnVersion{5b1ee2f-D}\def\svnDate{2020/07/01}\def\cmsCernNoTag{CERN-EP-2019-270}\def\cmsCernDate{\today}\def\cmsMessage{"Published in the Journal of High Energy Physics as \href{http://dx.doi.org/10.1007/JHEP06(2020)146}{\doi{10.1007/JHEP06(2020)146}.}"}

\begin{document}\cmsNoteHeader{TOP-15-018}

\hyphenation{had-ron-i-za-tion}
\hyphenation{cal-or-i-me-ter}
\hyphenation{de-vices}
\newcommand{\glu}{{\Pg}{\Pg}}
\newcommand{\qg}{{\Pq}{\Pg}}
\newcommand{\aofb}{\ensuremath{A^{(1)}_\mathrm{FB}}}
\newcommand{\pard}{\ensuremath{\hat{d}_{\PQt}}}
\newcommand{\parmu}{\ensuremath{\hat{\mu}_{\PQt}}}
\newcommand{\ac}{\ensuremath{A_{\mathrm{C}}}}
\newcommand{\cstar}{\ensuremath{c^{*}}}
\newcommand{\xF}{\ensuremath{x_{\text{F}}}}
\newcommand{\mtt}{\ensuremath{m_{\ttbar}}}
\newcommand{\cstarsq}{\ensuremath{c^{*2}}}
\newcommand{\axF}{\ensuremath{\abs{\xF}}}
\newcommand{\pL}{\ensuremath{p_{\mathrm{L}}}}
\newcommand{\csubr}{\ensuremath{c^{*}_{\mathrm{r}}}}
\newcommand{\xr}{\ensuremath{x_{\mathrm{r}}}}
\newcommand{\mr}{\ensuremath{m_{\mathrm{r}}}}
\newcommand{\muR}{\ensuremath{\mu_{\mathrm{R}}}}
\newcommand{\muF}{\ensuremath{\mu_{\mathrm{F}}}}

\cmsNoteHeader{TOP-15-018}

\title{Measurement of the top quark forward-backward production asymmetry and the anomalous chromoelectric and chromomagnetic moments in {\Pp}{\Pp} collisions at $\sqrt{s}=13\TeV$}

\date{\today}

\abstract{
   The parton-level top quark ({\PQt}) forward-backward asymmetry and the anomalous chromoelectric (\pard) and chromomagnetic (\parmu) moments have been measured using LHC {\Pp}{\Pp} collisions at a center-of-mass energy of 13\TeV, collected in the CMS detector in a data sample corresponding to an integrated luminosity of 35.9\fbinv. The linearized variable \aofb\ is used to approximate the asymmetry. Candidate \ttbar events decaying to a muon or electron and jets in final states with low and high Lorentz boosts are selected and reconstructed using a fit of the kinematic distributions of the decay products to those expected for \ttbar final states. The values found for the parameters are $\aofb=0.048^{+0.095}_{-0.087}\stat^{+0.020}_{-0.029}\syst$, $\parmu=-0.024^{+0.013}_{-0.009}\stat^{+0.016}_{-0.011}\syst$, and a limit is placed on the magnitude of $\abs{\pard}<0.03$ at 95\% confidence level.
}

\hypersetup{
pdfauthor={CMS Collaboration},
pdftitle={Measurement of the top quark forward-backward production asymmetry and the anomalous chromoelectric and chromomagnetic moments in pp collisions at sqrt(s)=13 TeV},
pdfsubject={CMS},
pdfkeywords={CMS, physics, analysis, top}}

\maketitle

\section{Introduction}
\label{sec:introduction}

The top quark (\PQt) is the most massive of the known fundamental particles, with a mass ($m_{\PQt}$) that is close to the electroweak scale.  The Yukawa coupling of the top quark to the Higgs field is close to unity~\cite{PDG2018}, which suggests that the top quark may play a role in electroweak symmetry breaking.  The top quark is also the only color-triplet fermion that decays before forming color-singlet bound states.  This provides a way to study its fundamental interactions with gauge bosons without the complications caused by hadronization.

\begin{figure}[hbt]
  \centering
    \includegraphics[width=\linewidth]{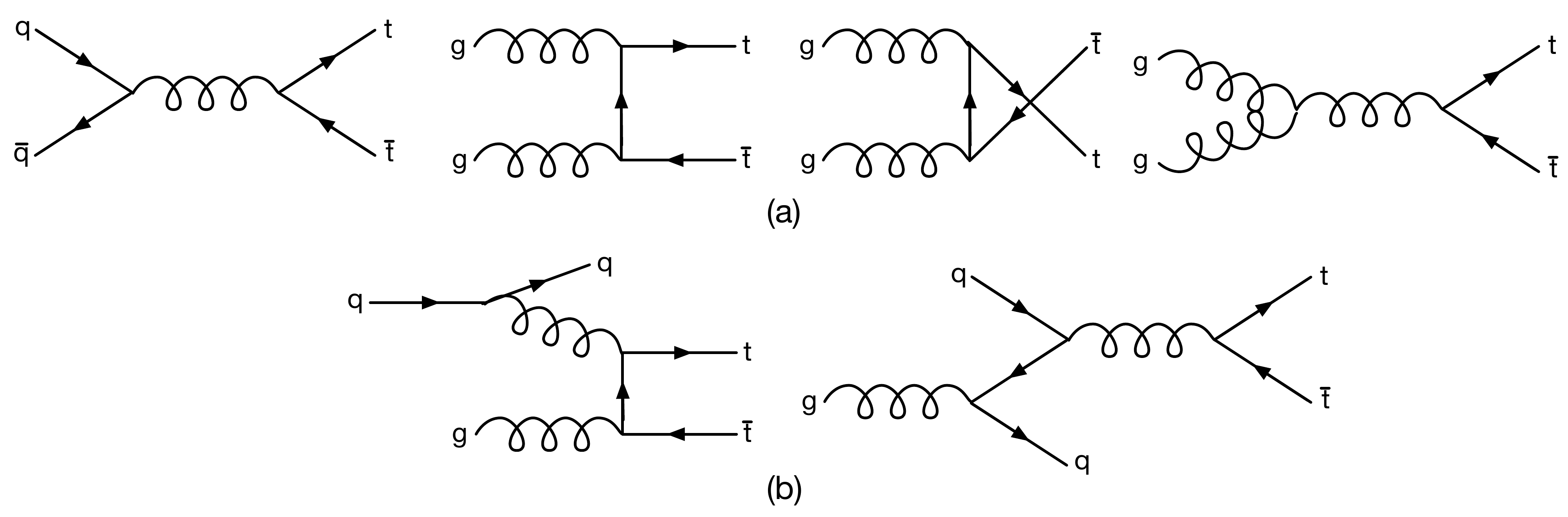}
  \caption{(a) Feynman diagrams for LO {\qqbar}- and {\glu}-initiated \ttbar subprocesses, and (b) example diagrams for the NLO {\qg}-initiated subprocess. }
    \label{fig:feynman}
\end{figure}

At the CERN LHC, according to the standard model (SM) of particle physics, top quarks are produced predominantly in pairs via the strong interaction, as described by quantum chromodynamics (QCD).  Feynman diagrams for leading-order (LO) quark-antiquark (\qqbar) and gluon-gluon (\glu) initiated subprocesses are shown in Fig.~\ref{fig:feynman}(a), and example diagrams for next-to-leading-order (NLO) quark-gluon (\qg) initiated subprocess are shown in Fig.~\ref{fig:feynman}(b). The NLO QCD calculations predict approximately 6\% \qqbar\ and 69\% \glu\ production of top quark pairs at a center-of-mass energy of 13\TeV.

\begin{figure}[hbt]
  \centering
    \includegraphics[width=0.45\linewidth]{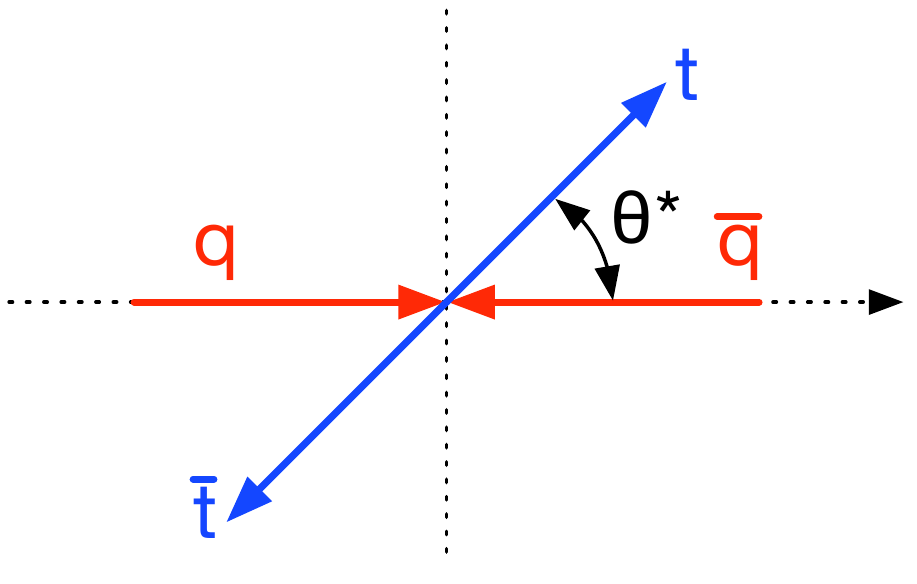}
  \caption{ Production angle $\theta^*$ in a $\qqbar \to \ttbar$ event defined in the \ttbar center-of-mass frame.}
    \label{fig:thetastar_angle_diagram}
\end{figure}

We describe here a search for anomalies in the angular distribution of produced \ttbar\ pairs. Those anomalies can be caused by modifications of the top quark-antiquark-gluon ({\ttbar}{\Pg}) vertex or by the presence of heavy states coupled to top quarks~\cite{Djouadi:2009nb, Frampton:2009rk, Ferrario:2009ee, Jung:2009pi, Cao:2009uz, Barger:2010mw, Cao:2010zb, Bauer:2010iq, Alvarez:2010js, Chen:2010hm, Bai:2011ed, Zerwekh:2011wf, Gresham:2011pa}. They are characterized through their impacts on the distribution of $\cstar=\cos\theta^*$, where $\theta^*$ is the production angle of the top quark relative to the direction of the initial-state parton in the \ttbar center-of-mass frame, as shown in Fig.~\ref{fig:thetastar_angle_diagram} for a $\qqbar \to \ttbar$ event.

When subprocesses contain an initial-state quark or antiquark, the sign of \cstar\ follows from the relative directions of the initial-state quark and the top quark (or the initial-state antiquark and the top antiquark).  We search separately for anomalies in the \cstar-antisymmetric (linearly dependent on \cstar) and the \cstar-symmetric (dependent only on \cstarsq) distribution functions, using the former to measure the top quark forward-backward asymmetry (\AFB) and the latter to measure the anomalous chromoelectric (\pard) and chromomagnetic (\parmu) dipole moments of the \ttbar\ vertex.

The parton-level forward-backward asymmetry is defined as
\begin{linenomath}\begin{equation}
 \AFB = \frac{\sigma(\cstar>0) - \sigma(\cstar<0)}{\sigma(\cstar>0) + \sigma(\cstar<0)} ,
 \label{AFBdef}
\end{equation}\end{linenomath}
where $\sigma(\cstar>0)$ and $\sigma(\cstar<0)$ represent the cross sections for production of the top quark in the forward and backward hemispheres, respectively, relative to the incident quark direction. The NLO effects generate positive values for \AFB\ in \qqbar-initiated subprocesses~\cite{Kuhn:1998kw, Kuhn:2011ri, AguilarSaavedra:2012rx, Czakon:2014xsa}.

A related \AFB\ quantity was measured by the CDF and D0 experiments at the Fermilab Tevatron, a proton-antiproton collider that operated at a center-of-mass energy $\sqrt{s}=1.96$\TeV, where the \qqbar\ subprocess dominates \ttbar\ production.  Because the quantity measured at the Tevatron includes \glu\ and \qg\ initial states, it is expected to be smaller than the quantity defined in Eq.~(\ref{AFBdef}) by a factor of $\approx0.85$ (about 18\% smaller).  Initial measurements~\cite{Abazov:2007ab,Aaltonen:2011kc} were somewhat larger, especially at CDF, than expected at the NLO level; however, more recent results~\cite{Abazov:2014cca,Aaltonen:2016bqv,Aaltonen:2017efp} are consistent with the SM.  Previous LHC measurements sensitive to the top quark angular production asymmetry performed by the ATLAS~\cite{ATLAS:2012an,Aad:2015lgx} and CMS~\cite{Chatrchyan:2011hk,Chatrchyan:2014yta,Khachatryan:2015mna,Khachatryan:2016ysn} Collaborations focused on the top quark charge asymmetry {\ac}, which does not separate the \qqbar\ initial states from the \glu\ and \qg\ initial states, and therefore uses only part of the available information.

This work represents a different approach that adopts a simplified model for the production mechanism using a likelihood analysis to separate the \qqbar\ subprocess from the \glu\ and \qg\ subprocesses as well as from other backgrounds.  The model provides an LO description of several possible ``beyond the SM processes''~\cite{Djouadi:2009nb, Frampton:2009rk, Ferrario:2009ee, Jung:2009pi, Cao:2009uz, Barger:2010mw, Cao:2010zb, Bauer:2010iq, Alvarez:2010js, Chen:2010hm, Bai:2011ed, Zerwekh:2011wf, Gresham:2011pa}, and a reasonable approximation for expected NLO QCD effects~\cite{Kuhn:1998kw}.  The differential cross section for $\qqbar \to \ttbar$ can be expressed as a linear combination of symmetric and antisymmetric functions of the production angle, with the antisymmetric function approximated as a linear function of \cstar:
\begin{linenomath}\begin{equation}
\frac{\rd\sigma}{\rd\cstar}(\qqbar) \approx f_{\text{sym}}(\cstar) + \left[ \int_{-1}^{1} f_{\text{sym}}(x) \rd{}x \right] \cstar \aofb(\mtt).
\label{Aofbdef}
\end{equation}\end{linenomath}
The symmetric function $f_{\text{sym}}$ depends only on kinematic properties of the event and \aofb\ is a parameter that depends upon \mtt, the invariant mass of the \ttbar\ system.  Using Eq.~(\ref{Aofbdef}) to evaluate Eq.~(\ref{AFBdef}), we find that $\AFB \approx \aofb$, which defines the linearized forward-backward asymmetry. Equation~(\ref{Aofbdef}) describes the LO exchange and interference terms expected from $s$-channel resonances with chiral couplings.  Because the statistical power of the \ttbar\ sample is not sufficient to measure \aofb\ as a function of \mtt, we measure only an average value over the entire sample.  This approximation is similar to the mass-averaged leading term of a Legendre polynomial used by the CDF Collaboration in 2013~\cite{CDF:2013gna} to characterize the angular distribution of their \ttbar\ data.  

The validity of this approximation is verified by fitting the linearized function given in Eq.~(\ref{eq:qqonedef}) to data simulated with the NLO \POWHEG\ generator~\cite{Frixione:2007nw} and comparing the resulting \aofb\ values with the $\AFB$ values determined from counting events with positive and negative \cstar.  The results of this study are presented in Table~\ref{tab:afbcomp}.  Samples of simulated {\Pp}{\PAp} and {\Pp}{\Pp} events at $\sqrt{s}=1.96\TeV$ and $\sqrt{s}=13\TeV$, respectively, are also subdivided into two \mtt\ regions to perform the analysis.  Comparing the results in the full samples and in the \mtt\ regions, we conclude that the linear approximation is accurate to within a few percent of measured asymmetries and that it reflects the average asymmetry of each sample.  We further note that smaller values of the {\Pp}{\Pp} asymmetry at $\sqrt{s}=13\TeV$ than the {\Pp}{\PAp} asymmetry at $\sqrt{s}=1.96\TeV$ appear to be due to the more copious production of events with energetic extra jets which contribute negatively to the average asymmetry.  It is clear that our LHC measurements are not directly comparable with those from the Tevatron experiments.

\begin{table}[htb]
\topcaption{The accuracy of the linear approximation used to define \aofb\ is assessed by comparing the results of linear fitting to determine \aofb\ and of event counting to determine $\AFB$.  The  {\Pp}{\PAp} and {\Pp}{\Pp} data samples are simulated with the NLO \POWHEG\ generator~\cite{Frixione:2007nw} at center-of-mass (CM) energies of 1.96\TeV\ and 13\TeV, respectively.}
\centering
\begin{tabular}{c c c c c}
Initial State & CM Energy & Mass Range       & $\AFB$ (counting) & $\aofb$ (fitting) \\
\hline
 {\Pp}{\PAp}  & 1.96\TeV  & $\mtt<500\GeV$   & 0.0535$\pm$0.0012 & 0.0546$\pm$0.0011 \\
 {\Pp}{\PAp}  & 1.96\TeV  & $\mtt>500\GeV$   & 0.0998$\pm$0.0024 & 0.1044$\pm$0.0022 \\
 {\Pp}{\PAp}  & 1.96\TeV  & all              & 0.0626$\pm$0.0011 & 0.0639$\pm$0.0010 \\
 \hline
 {\Pp}{\Pp}   & 13\TeV    & $\mtt<500\GeV$   & 0.0249$\pm$0.0016 & 0.0251$\pm$0.0014 \\ 
 {\Pp}{\Pp}   & 13\TeV    & $\mtt>500\GeV$   & 0.0414$\pm$0.0022 & 0.0414$\pm$0.0020 \\ 
 {\Pp}{\Pp}   & 13\TeV    & all              & 0.0306$\pm$0.0013 & 0.0305$\pm$0.0012 \\ 
\hline
\end{tabular}
\label{tab:afbcomp}
\end{table}

Several authors have considered the impact of possible top quark anomalous chromoelectric and chromomagnetic dipole moments on the production of a \ttbar\ system at hadron colliders~\cite{Atwood:1992vj,Atwood:1994vm,Haberl:1995ek,Cheung:1995nt}.  We follow the conventions and results of Ref.~\cite{Haberl:1995ek}, and define \pard\ and \parmu\ in terms of an effective Lagrangian of the kind
\begin{linenomath}\begin{equation}
{\cal L}_{\ttbar\Pg}=-g_s\,\left[\PAQt \gamma^\mu G_\mu \PQt + i\frac{\pard}{2m_{\PQt}} \; \PAQt \sigma^{\mu\nu}\gamma_5G_{\mu\nu} \PQt
+\frac{\parmu}{2m_{\PQt}} \;  \PAQt \sigma^{\mu\nu}G_{\mu\nu}\PQt\right] ,
\label{Lttg}
\end{equation}\end{linenomath}
where $\sigma^{\mu\nu}=\frac{i}{2}[\gamma^\mu,\gamma^\nu]$,
$G_\mu=G_\mu^aT^a$; where the $G_\mu^a$ are the gluon fields and the
$SU(3)_C$ generators are $T^a=\frac{1}{2}\lambda^a$ ($a$=1, \ldots, 8);
the $G_{\mu\nu}=G_{\mu\nu}^aT^a$, with the gluon field-strength
tensors being $G_{\mu\nu}^a=\partial_\mu G_\nu^a-\partial_\nu G_\mu^a
-g_sf_{abc}G_\mu^bG_\nu^c$, and the $f_{abc}$ being $SU(3)$ group structure constants.

This paper discusses our measurements of \aofb, \pard, and \parmu\ in {\Pp}{\Pp} collisions at the CERN LHC.  The analysis uses theoretical models of the \ttbar\ cross section as functions of \aofb, \pard, and \parmu\ to describe the angular distributions observed in a sample of \ttbar-enriched events collected at the CMS experiment.  The analysis is based on final states containing a single lepton (muon or electron) and several jets, usually referred to as lepton+jets events. Our measurement of the \aofb\ of top quarks in initial \qqbar\ states is the first of its kind at the LHC.  Values of \pard\ and \parmu\ have previously been extracted from \ttbar spin correlations~\cite{Khachatryan:2016xws}, but this is the first measurement that relies on differential distributions in \ttbar events.

This paper is organized as follows. Section~\ref{sec:analysis_strategy} details the strategies used in extracting parameters of interest from the observed cross section. Section~\ref{sec:detector} briefly describes the CMS detector and the final-state objects used in the analysis. Section~\ref{sec:samples} discusses the recorded and simulated events. Section~\ref{sec:selection} describes the event selection and kinematic reconstruction of the \ttbar events. Section~\ref{sec:background} discusses the fitting method and the techniques used to estimate the backgrounds. Section~\ref{sec:systematics} describes the systematic uncertainties associated with the analysis. Section~\ref{sec:results} reviews the results of the fits to the data, and Section~\ref{sec:summary} provides a brief summary.

\section{Analysis strategy}
\label{sec:analysis_strategy}

Measuring the top quark forward-backward asymmetry at the LHC is considerably more challenging than at the Tevatron, where the \ttbar cross section is dominated by the \qqbar\ subprocess and the incident quark and antiquark directions are well defined by the proton and antiproton beams.  At the LHC, however, \ttbar production is dominated by the \glu\ subprocess, from which no asymmetry arises, thus complicating the extraction of the asymmetry in the $\qqbar \to \ttbar$ subprocess.  To separate the \qqbar\ from the \glu\ and \qg\ subprocesses, and from other backgrounds, we use the quantities \mtt, \xF, and \cstar\ to describe \ttbar events, where: $\xF = 2\pL/\sqrt{s}$ is the scaled longitudinal momentum \pL\ of the \ttbar system in the laboratory frame; and \cstar, \mtt\ were defined in Section~\ref{sec:introduction}. The use of \xF\ to separate events with different initial states is a technique that is well-established in the literature, and its application to a measurement of \AFB\ in top quark pair production was proposed in~\cite{AguilarSaavedra:2012va}.

\begin{figure}[hbt]
  \centering
    \includegraphics[width=0.44\linewidth]{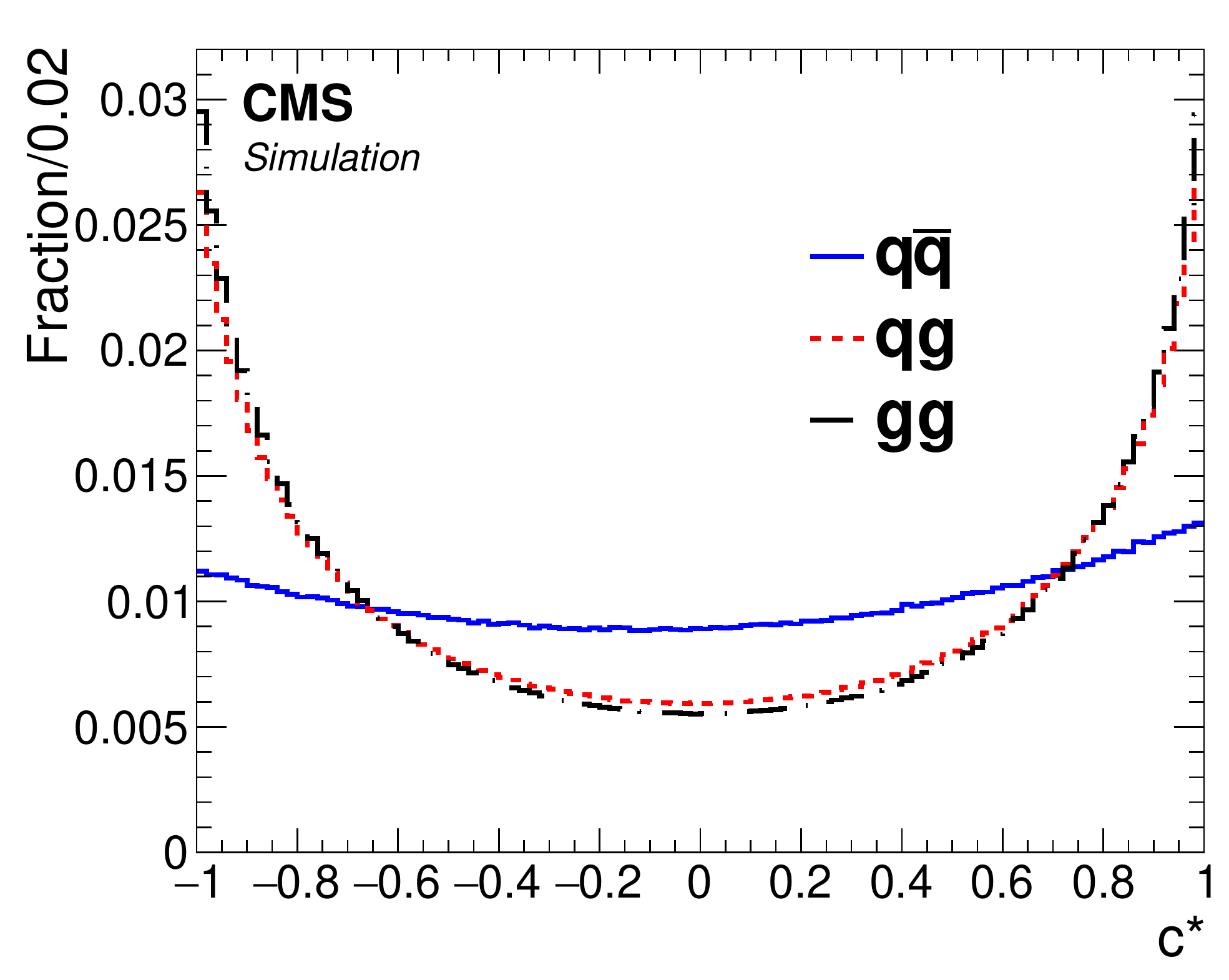}
    \includegraphics[width=0.44\linewidth]{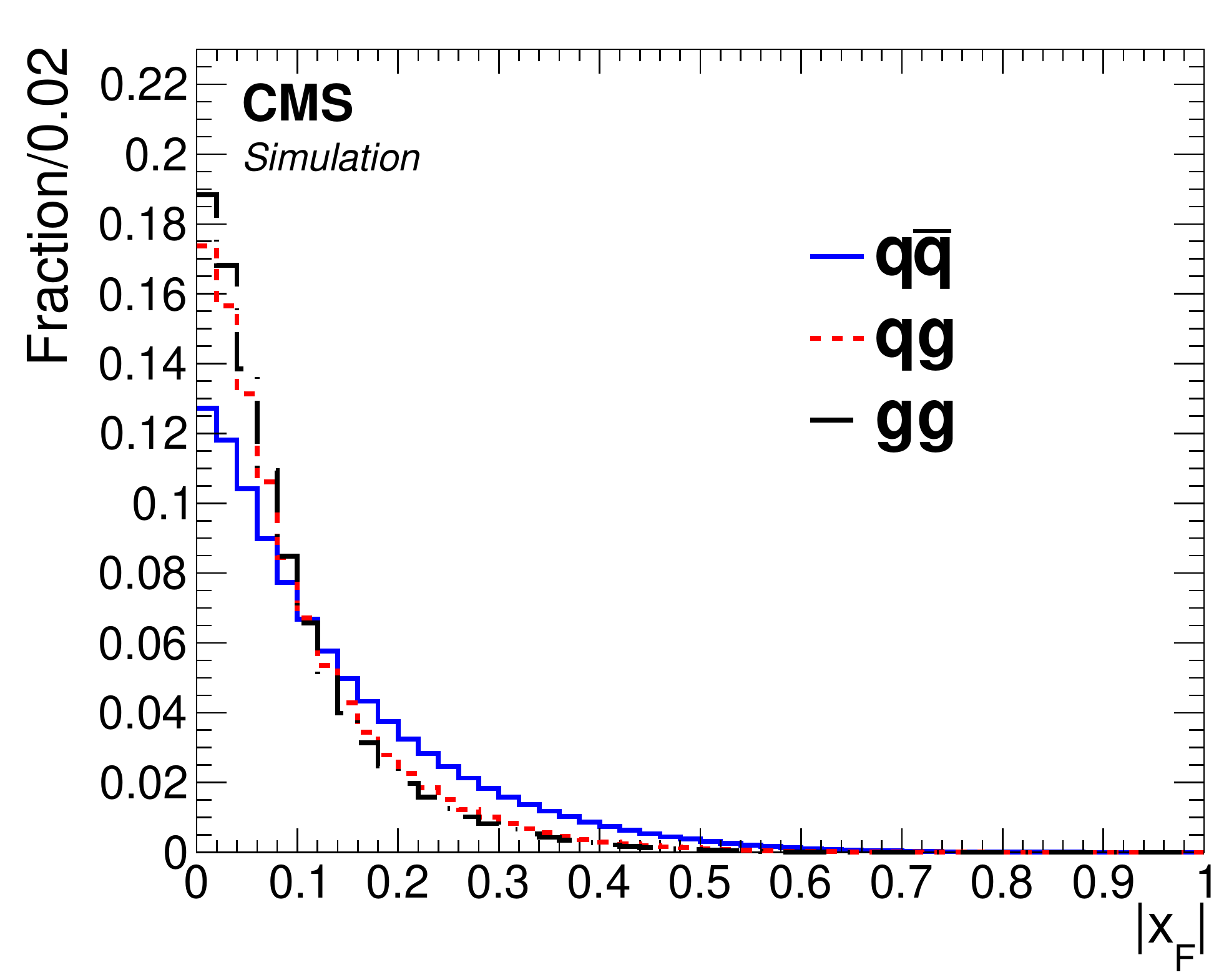}
    \includegraphics[width=0.44\linewidth]{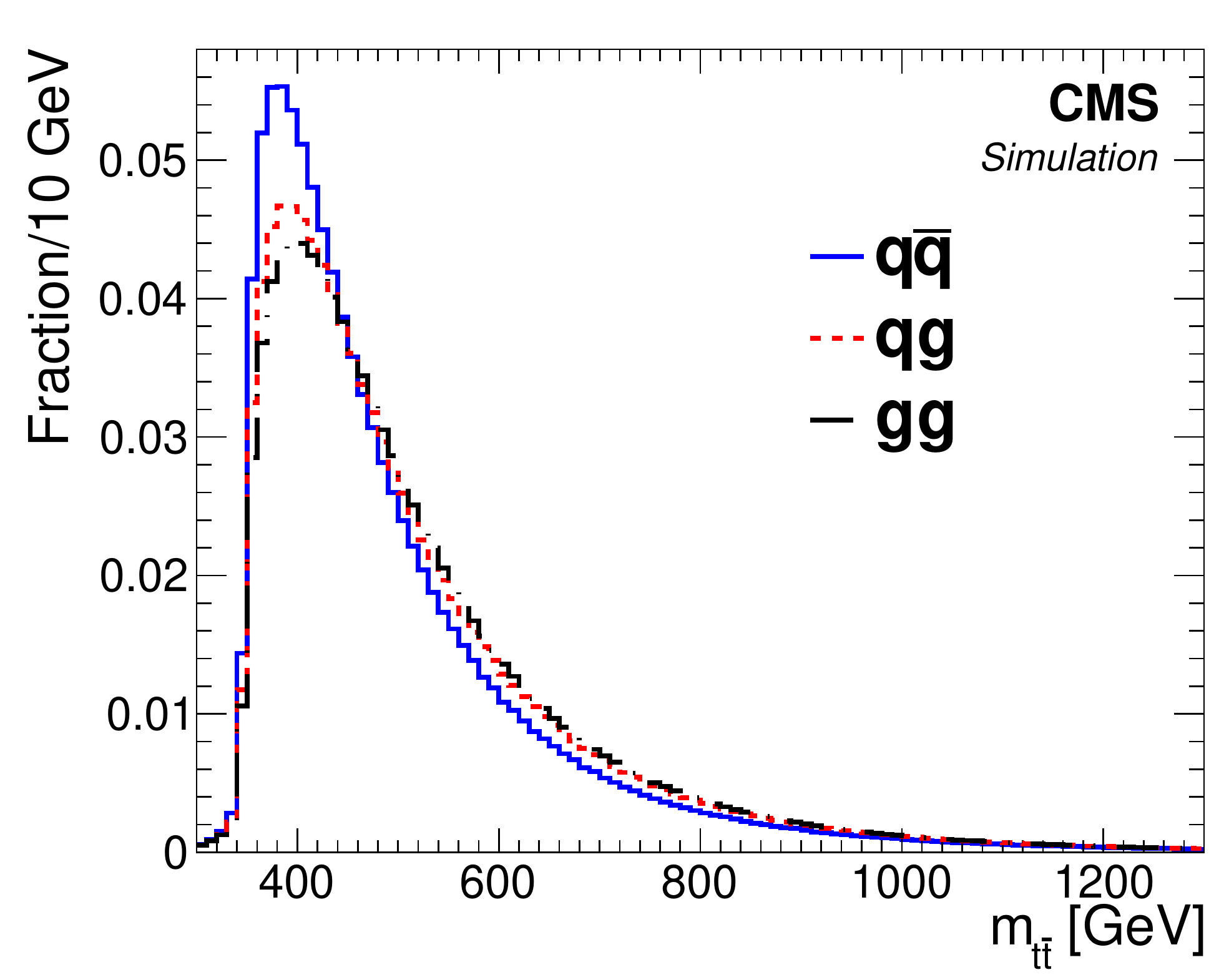}
    \includegraphics[width=0.44\linewidth]{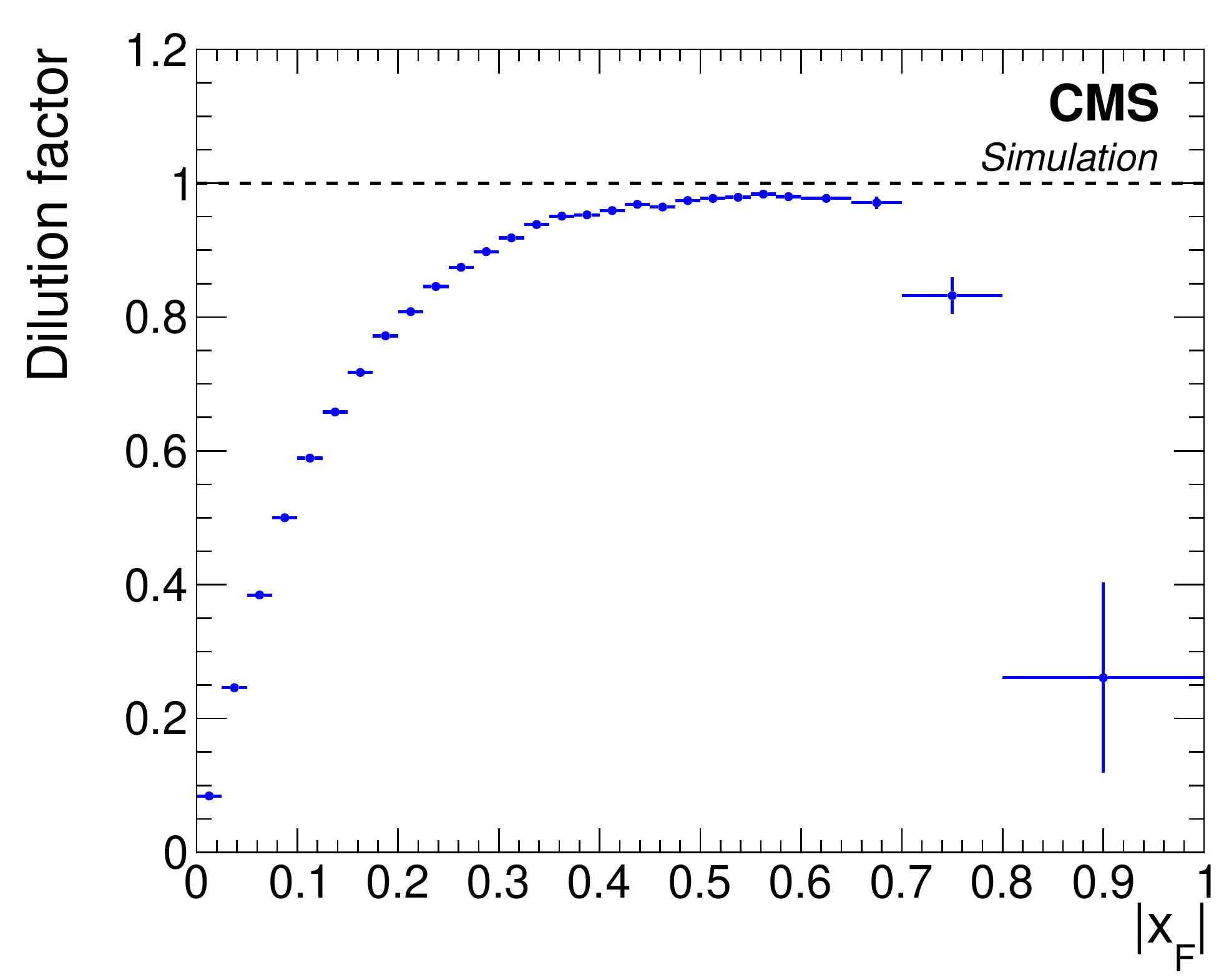}
  \caption{The generator-level \cstar\ (upper left), \axF\ (upper right), and \mtt\ (lower left) distributions normalized to unity for the subprocesses \qqbar, \qg, and $\glu \to \ttbar$.  The dilution factor, when taking the longitudinal direction of the \ttbar pair in the lab frame as the quark direction for \qqbar\ events, is shown in the lower right plot as a function of \axF.}
    \label{fig:distributions}
\end{figure}

The distributions in \mtt, \xF, and \cstar\ for the \glu, \qg, and \qqbar\ initial states for \ttbar events simulated using the NLO \POWHEG\ v2~\cite{Nason:2004rx,Frixione:2007vw,Alioli:2010xd} Monte Carlo (MC) generator for {\Pp}{\Pp}\ collisions at $\sqrt{s}=13\TeV$ are shown in Fig.~\ref{fig:distributions}; the \glu\ and \qg\ distributions are quite similar.  Because of this similarity, and because the SM asymmetry for \qg\ events is expected to be approximately two orders of magnitude smaller than for \qqbar\ events~\cite{Kuhn:1998kw}, the \glu\ and \qg\ subprocesses are combined into a single distribution function in our analysis.  The \mtt\ distribution for \qqbar\ events is somewhat narrower than for the other processes. The \cstar\ distribution for \qqbar\ events is more isotropic than that for the \glu\ and \qg\ processes due to the $t$-channel pole that dominates their cross sections.  Of key importance is that the \xF\ distribution for the \qqbar\ events has a longer tail that helps to discriminate them and to correctly identify the incident-quark direction.  Because the gluon and antiquark parton distribution functions (PDFs) of the proton have a lower average transverse momentum than those of the quark parton, the direction of the \ttbar system \pL\ is correlated with the initial quark direction in {\qqbar}-initiated events.  This provides a reasonable choice for the initial parton direction as the direction of the Collins--Soper frame~\cite{Collins:1977iv}. The result of taking the longitudinal direction of the \ttbar pair in the lab frame as the quark direction is shown in Fig.~\ref{fig:distributions}: defining $N_{\mathrm{C}}$ as the number of correct assignments and $N_{\mathrm{I}}$ as the number of incorrect assignments, the dilution factor $D=(N_{\mathrm{C}}-N_{\mathrm{I}})/(N_{\mathrm{C}}+N_{\mathrm{I}})$ becomes large in the \qqbar\ enriched region of large \axF.  The reduced $D$ in the largest \axF\ bins appears to be an artifact of the NNPDF3.0 PDF set~\cite{Ball:2014uwa} used to generate the plot, and it does not affect this analysis as no data events are observed with such large values of \axF.

The differential cross section for $\qqbar \to \ttbar$ can be written in the center-of-mass frame as
\begin{linenomath}\begin{equation}
\frac{\rd\sigma}{\rd\cstar}(\qqbar) = K\frac{\pi\alpS^2}{9\mtt^2}\beta\left\lbrace 2-\beta^2+\beta^2\cstarsq+\alpha\left(1-\beta^2\cstarsq\right)
+2\left[2-\frac{2}{3}\beta^2+\alpha\left(1-\frac{1}{3}\beta^2\right)\right]\aofb\cstar\right\rbrace ,
\label{eq:qqonedef}
\end{equation}\end{linenomath}
where $K$ is an NLO normalization factor, $\alpS = g_s^2/4\pi$ is the strong coupling constant, $\beta=\sqrt{\smash[b]{1-4m_{\PQt}^2/\mtt^2}}$ is the velocity of the top quark in the center-of-mass frame, and $\alpha(\beta)$ is the longitudinal polarization of the exchanged gluon. This parameterization describes the process to NLO precision, using a linear approximation for the NLO forward-backward asymmetry that is also an LO description of the effects of possible $s$-channel vector resonances in extensions of the standard model~\cite{Cao:2010zb,Gresham:2011pa}.

At LO, the presence of the \pard\ and \parmu\ terms modify both the \qqbar\ and \glu\ contributions to the  {\Pp}{\Pp}\ $\to$ \ttbar cross section.  As in the case of \aofb, we use a framework that approximates SM NLO contributions with good accuracy, and provides possible anomalous contributions at LO.  The \qqbar\ cross section can then be expressed as follows~\cite{Haberl:1995ek}:
\begin{linenomath}\begin{equation}\begin{aligned}
\frac{\rd\sigma}{\rd\cstar}(\qqbar) &= K\frac{\pi\alpS^2}{9\mtt^2}\beta\biggl\lbrace 2 - \beta^2+\beta^2c^{*2}+\alpha\left(1-\beta^2\cstarsq\right) \\
&+4(2\parmu+\parmu^2-\pard^2)+4\left(\parmu^2+\pard^2\right)\frac{1-\beta^2\cstarsq}{1-\beta^2} \biggr\rbrace ,
\label{eq:qqtwodef}
\end{aligned}\end{equation}\end{linenomath}
where the NLO asymmetry described in Eq.~(\ref{eq:qqonedef}) has been suppressed because the anomalous moments affect only the symmetric part of the cross section.  Note that the last term is strongly enhanced at large values of \mtt\ (as $\beta\to1$).  The \glu\ cross section can be expressed as~\cite{Haberl:1995ek}:
\begin{linenomath}\begin{equation}\begin{aligned}
\frac{\rd\sigma}{\rd\cstar}(\glu) = K\frac{\pi\alpS^2\beta}{48\mtt^2}\biggl\lbrace & \frac{7+9\beta^2\cstarsq}{1-\beta^2\cstarsq}\left[\frac{1-\beta^4 c^{*4}+2\beta^2(1-\beta^2)(1-\cstarsq)}{2(1-\beta^2\cstarsq)} \left(1+\varepsilon \beta^2\cstarsq\right)+\parmu(1+\parmu)\right] \nonumber \\
&+8(\parmu^2+\pard^2)\left(\frac{7(1+\parmu)}{1-\beta^2}+\frac{1-5\parmu}{2(1-\beta^2\cstarsq)}\right) \nonumber \\
&+8(\parmu^2+\pard^2)^2\left(\frac{1}{1-\beta^2\cstarsq}+\frac{1}{1-\beta^2}+\frac{4(1-\beta^2\cstarsq)}{(1-\beta^2)^2}\right)\biggr\rbrace ,
\label{eq:ggonedef}
\end{aligned}\end{equation}\end{linenomath}
where the SM NLO contributions are parameterized through the factor $K$ and the empirical $\varepsilon=\varepsilon(\beta)$ function.  As for \qqbar, the effects from anomalous moments are most significant as $\beta\to1$.

Because the effects from as yet undiscovered massive particles and from the anomalous moments \pard\ and \parmu\ are most noticeable at large \mtt, and because the fraction of \qqbar\ events increases with top quark-pair momentum, the first of the three \ttbar decay topologies we consider, called ``type-1,'' comprises events with high Lorentz boost in which the decay products of the hadronic top quark are fully merged into a single jet.  The second topology, called ``type-2,'' consists of events that have a large, high-momentum, high-mass jet that indicates either a partially- or fully-merged hadronic top quark decay, but which lack a single large jet definitively identified as originating from a merged top quark decay. These type-2 events are included to bridge the gap between events where the top quark decay products are fully merged or fully resolved.  The type-1 and -2 topologies are collectively referred to as ``boosted'' because their decaying top quarks have high \pt. The third and most populated category, called ``type-3,'' includes events with low-mass jets; type-3 events are also called ``resolved'' events because all decay products are individually distinguishable. The fully selected event sample comprises approximately 2.2\% type-1 (7195 {\Pgm}+jets and 3108 {\Pe}+jets), 11.6\% type-2 (50311 {\Pgm}+jets and 3735 {\Pe}+jets), and 86.2\% type-3 (234839 {\Pgm}+jets and 166213 {\Pe}+jets events) events.

As detailed in Section~\ref{sec:background}, linear combinations of normalized three-dimensional distributions (templates) in \mtt, \xF, and \cstar\ reconstructed from simulated SM \ttbar events generated at NLO are used in a simultaneous likelihood fit to the observed differential cross section in 12 total channels defined by the decay topology, lepton charge, and lepton flavor. Generator-level information is used to separate \qqbar\ from \qg\ and \glu\ events and to reweight them using Eqs.~(\ref{eq:qqonedef})--(\ref{eq:ggonedef}) to produce the parameter-independent templates. Other MC contributions and events in data are used to construct templates representing the background. A total general linear combination of these templates is fitted to data to independently extract the three values of \aofb, \pard, and \parmu.

\section{The CMS detector and physics objects}
\label{sec:detector}

The central feature of the CMS apparatus is a superconducting solenoid of 6\unit{m} internal diameter, providing a magnetic field of 3.8\unit{T}. Within the solenoid volume are a silicon pixel and strip tracker, a lead tungstate crystal electromagnetic calorimeter (ECAL), and a brass and scintillator hadron calorimeter (HCAL), each composed of a barrel and two endcap sections. Forward calorimeters extend the pseudorapidity coverage provided by the barrel and endcap detectors. Muons are detected in gas-ionization chambers embedded in the steel flux-return yoke outside the solenoid. A more detailed description of the CMS detector, together with a definition of the coordinate system used and the relevant kinematic variables, can be found in Ref.~\cite{Chatrchyan:2008zzk}.

Events of interest are selected using a two-tiered trigger system~\cite{Khachatryan:2016bia}. The first level, composed of custom hardware processors, uses information from the calorimeters and muon detectors to select events at a rate of around 100\unit{kHz} within a time interval of less than 4\mus. The second level, known as the high-level trigger, consists of a farm of processors running a version of the full event reconstruction software optimized for fast processing, and reduces the event rate to around 1\unit{kHz} before data storage.

The particle-flow (PF) algorithm~\cite{CMS-PRF-14-001} aims to reconstruct and identify each individual particle in an event using an optimized combination of information from the various elements of the CMS detector. Muons are identified in the range $\abs{\eta} < 2.4$, with detection planes made using three technologies: drift tubes, cathode strip chambers, and resistive plate chambers. The energy of each muon is obtained from the curvature of the corresponding track. The energy of each electron is determined from a combination of the electron momentum at the primary interaction vertex (PV) as determined by the tracker, the energy of the corresponding ECAL cluster, and the energy sum of all bremsstrahlung photons spatially compatible with originating from the electron track. The energy of each charged hadron is determined from a combination of its momentum measured in the tracker and the matching ECAL and HCAL energy deposits, corrected for zero-suppression effects and for the response function of the calorimeters to hadronic showers. Finally, the energy of each neutral hadron is obtained from the corresponding corrected ECAL and HCAL energy.

The candidate vertex with the largest value of summed physics-object $\pt^2$ is taken to be the primary $\Pp\Pp$ interaction vertex. The physics objects are the jets, clustered using the anti-\kt\ (AK) jet finding algorithm~\cite{Cacciari:2008gp,Cacciari:2011ma} with the tracks assigned to candidate vertices as inputs, and the associated missing transverse momentum. The missing transverse momentum vector \ptvecmiss\ in an event is defined as the negative vector sum of the transverse momenta of all the reconstructed PF objects. Its magnitude is denoted as \ptmiss~\cite{Sirunyan:2019kia}. The \ptvecmiss\ is modified to account for corrections to the energy scale of the reconstructed jets in the event. Anomalous high-\ptmiss\ events can be due to a variety of reconstruction failures, detector malfunctions or non collisions backgrounds. Such events are rejected by event filters that are designed to identify more than 85--90\% of the spurious high-\ptmiss\ events with a misidentification rate less than 0.1\%~\cite{Sirunyan:2019kia}.

Jets called ``AK4" (``AK8") are clustered from the PF objects in an event using the anti-\kt\ algorithm with a distance parameter of 0.4 (0.8). Jet momentum is determined as the vectorial sum of all PF object momenta in the jet, and is found to be, on average, within 5 to 10\% of the true momentum over the whole \pt\ spectrum and detector acceptance, based on simulation.  Additional {\Pp}{\Pp} interactions within the same or nearby bunch crossings (pileup) can contribute extra tracks and calorimetric energy depositions, increasing the apparent jet momentum. To mitigate this effect, tracks originating from pileup vertices are discarded, and an offset correction is applied to account for remaining contributions. Jet energy corrections are derived from simulation studies~\cite{Khachatryan:2016kdb} so the average measured response of jets becomes identical to that of particle-level jets. In situ measurements of the momentum balance in dijet, photon+jet, {\PZ}+jet, and multijet events are used to detect any residual differences between the jet energy scale in data and in simulation, and appropriate corrections are made. Additional selection criteria are applied to each jet to identify jets potentially dominated by instrumental effects or reconstruction failures~\cite{CMS-PAS-JME-10-003}.

To prevent double counting of lepton momentum in high-\pt\ (``boosted") leptonic top quark decays for which the final state lepton and \PQb\ quark jet are not expected to be well separated, leptons reconstructed and identified within jets are not taken into account in the jet momentum computation. Muons are removed from jets using a key-based algorithm, wherein a muon four-vector is subtracted from that of a reconstructed jet if the muon's PF candidate appears in the list of jet PF candidates, and electrons are removed from jets by subtracting their four-vectors from those of any reconstructed jets within {\DR} equal to the distance parameter of the jet clustering algorithm.

The ``soft drop'' (SD) algorithm~\cite{Dasgupta:2013ihk} with angular exponent $\beta = 0$, soft cutoff threshold $z_{\text{cut}} < 0.1$, and characteristic radius $R_{\mathrm{0}} = 0.8$~\cite{Larkoski:2014wba}, is applied to AK8 jets to remove soft, wide-angle radiation. In this algorithm the constituents of the AK8 jets are reclustered using the Cambridge--Aachen algorithm~\cite{Dokshitzer:1997in,Wobisch:1998wt}. Application of the SD algorithm improves discrimination between signal events, which contain jets from {\PW} boson or top quark decays, and background events, which consist of jets produced via the strong interaction, referred to as ``QCD multijet" events. This improves the jet mass resolution, reducing the mass of single light quark and gluon jets, and mitigating the effects of pileup~\cite{CMS:2017wyc}. The subjets identified by the algorithm are considered as individual objects themselves and are used in the reconstruction of the decaying top quark pair, as described in Section~\ref{sec:selection}.

\section{Data and simulation}
\label{sec:samples}

This analysis is performed using the data collected by the CMS detector in 2016 at $\sqrt{s}=13\TeV$, corresponding to an integrated luminosity of $35.9 \pm 0.9 \fbinv$~\cite{CMS-PAS-LUM-17-001}. Events in the type-1 and -2 channels pass a single-muon trigger requiring a muon with $\pt > 50 \GeV$, a single-electron trigger requiring an electron with $\pt > 115 \GeV$, or an electron+jets trigger with 50 and 165\GeV thresholds on the electron and the leading-jet \pt, respectively. Events in the type-3 channels must pass a single-muon trigger requiring an isolated muon with $\pt > 24\GeV$, or a single-electron trigger for electrons with $\pt > 27\GeV$. The average trigger efficiencies for events passing offline selection are approximately 95 (97)\% in the boosted (resolved) {\PGm}+jets channels and approximately 98 (74)\% in the boosted (resolved) {\Pe}+jets channels.

The \POWHEG\ MC generator~\cite{Frixione:2007nw} is used to simulate the \ttbar signal at NLO using the \PYTHIA\ 8.219 parton-shower generator~\cite{Sjostrand:2014zea} and the CUETP8M2T4 tune~\cite{CMS-PAS-TOP-16-021} assuming $m_{\PQt}=172.5 \GeV$. These MC events are also reweighted~\cite{Kalogeropoulos:2018cke} to model the dependence of the \ttbar distributions on renormalization and factorization scales $\muR=\muF=\sqrt{\smash[b]{m_{\PQt}^{2}+\pt^{2}}}$~\cite{Cacciari:2003fi,Catani:2003zt}, and the choice of the NNPDF3.0 PDFs~\cite{Ball:2014uwa}, with $\alpS=0.118$.

The \POWHEG\ and \PYTHIA\ generators using the same tune are also used to simulate contributions to the background from single top quark processes at NLO, both in the $t$ channel~\cite{Alioli:2009je} using \POWHEG\ and \textsc{madspin}~\cite{Artoisenet:2012st} and in the {\PQt}{\PW} channel with \POWHEG\ v1~\cite{Re:2010bp}. Background contributions from single top quark processes in the $s$ channel are simulated at NLO using the \MGvATNLO\ MC generator~\cite{Alwall:2014hca} matched to \PYTHIA\ parton showers; and contributions from Drell--Yan ({\PZ} or {\PGg}+jets) and {\PW}+jets production are simulated at LO using the \MADGRAPH\ 5 MC generator~\cite{Alwall:2014hca}, matched to \PYTHIA\ parton showers and the CUETP8M1 tune through the MLM prescription~\cite{Alwall:2007fs}.

All MC events are processed through a full simulation of the CMS detector in \GEANTfour\ \cite{Agostinelli:2002hh}. The distributions for MC events are normalized to their predicted cross sections at NLO for single top quark production in the $s$  and $t$ channels~\cite{Kant:2014oha}; at NLO using additional next-to-next-to-leading logarithms (NNLL) soft gluon correction for single top quark production in the {\PQt}{\PW} channel~\cite{Kidonakis:2010ux}; at next-to-next-to-leading order (NNLO) for {\PW}+jets and {\PZ}/{\PGg}+jets \cite{Gavin:2012sy,Gavin:2010az,Li:2012wna}; and at NNLO+NNLL for the \ttbar signal~\cite{Czakon:2011xx}. All MC samples are corrected to bring their generated pileup distributions into agreement with those observed in data. Identical selection criteria and reconstruction procedures are otherwise applied to all simulated events and data.

\section{Event reconstruction and selection}
\label{sec:selection}

The selection criteria are designed to identify \ttbar events in which one of the two top quarks decays to a charged lepton, a neutrino, and a single \PQb\ quark jet, and the other decays to only jets. The leptonically decaying top quark can yield a muon or electron in one of the three topologies, leading
to six mutually exclusive categories: $\Pgm+$jets and $\Pe+$jets, each of type-1, type-2, and type-3.

Muon (electron) candidates are required to have $\pt > 50$ (80)\GeV in the boosted topologies, and $\pt > 30\GeV$ in the resolved topology. Only lepton candidates in the range of $\abs{\eta} < 2.4$ are considered, and any electron candidates in the transition region between barrel and endcap calorimeters, corresponding to $1.44 < \abs{\eta} < 1.57$, are rejected. To suppress misidentified leptons resulting from the products of hadronization, accepted leptons are required to be isolated from nearby hadron activity in the event. Lepton isolation is determined in part with a ``2D selection'' applied as a logical ``or'' of two independent selections. The first is a requirement on the component of the lepton momentum that is transverse to the axis of the nearest AK4 jet, called $\pt^{\text{rel}}$, and the second is the distance $\DR = \sqrt{\smash[b]{(\Delta\eta)^2 + (\Delta\phi)^2}}$ with respect to the nearest AK4 jet. In both cases, the nearest AK4 jet must have $\pt > 15\GeV$ and $\abs{\eta} < 3.0$. To be considered isolated in the boosted topologies, a lepton must have $\pt^{\text{rel}} > 30\GeV$ or $\DR > 0.4$. In the resolved topology, lepton isolation is additionally determined according to the sum of the scalar \pt\ of the neutral and charged hadron PF candidates located within a cone of size $\DR = 0.4$ and $0.3$ for muons and electrons, respectively. This sum is required to be less than 15 (6)\% of the muon (electron) \pt, and only leptons passing both the 2D selection and these PF isolation criteria are considered isolated in the resolved topology. Any event containing more than one isolated lepton is rejected to suppress background from two-lepton \ttbar decays in which each top quark decays to leptons. Requirements are also imposed on the \ptmiss\ in each event to suppress background from multijet events containing a muon or electron. Events in the boosted muon+jets or electron+jets channels are required to have $\ptmiss > 50$ or 100\GeV, respectively, and resolved events of both lepton flavors are required to have $\ptmiss > 40\GeV$. Events failing \ptmiss\ requirements or lepton isolation requirements are used in the estimation of QCD multijet background contribution from data, as described in Section~\ref{sec:background}.

All AK4 and AK8 jets are required to have $\pt>30$ and $\pt>200\GeV$, respectively, and to be in the range of $\abs{\eta} < 2.4$. The AK8 jets are also required to have at least two subjets identified through the SD clustering algorithm. In type-1 events, at least one AK8 jet must be present and be identified (``tagged'') as originating from the merged decay of a top quark. These identified top quark (\PQt-tagged) jets are selected using simultaneous criteria on jet \pt; the jet mass after application of the SD algorithm; and the ungroomed $N$-subjettiness~\cite{Thaler:2010tr} substructure discriminant variable $\tau_{32}=\tau_{3}/\tau_{2}$, for which smaller values indicate a greater likelihood that the jet is composed of three rather than two subjets. An AK8 jet is considered \PQt-tagged if it has $\pt > 400\GeV$, an SD mass in the range $105 < m_{\text{AK8}}^{\text{SD}} < 220\GeV$, and $\tau_{32} < 0.81$. By contrast, type-2 events are required to have no \PQt-tagged AK8 jets, but must contain at least one AK8 jet with $m_{\text{AK8}}^{\text{SD}} > 40\GeV$, and at least four AK4 jets. Type-1 and -2 muon (electron) events require the highest momentum AK4 jet to have $\pt > 150$ (250)\GeV and the second-highest momentum AK4 jet to have $\pt > 50$ (70)\GeV. Type-3 events are required to contain zero selected AK8 jets, and at least four selected AK4 jets.

To help discriminate the \ttbar signal from {\PZ}/{\PGg}+jets and {\PW}+jets backgrounds, AK4 jets originating from decays of \PQb\ quarks are identified using an algorithm that combines lifetime information from tracks and secondary vertices~\cite{BTV-16-002}. Type-1 and -2 events are required to have at least one AK4 jet that is identified as a \PQb\ quark (\PQb-tagged) at the ``loose'' working point of the algorithm, which has an 83\% efficiency of correctly identifying a \PQb\ quark jet and a 9\% probability of misidentifying a gluon, \PQc\ quark, or light quark jet as a \PQb\ quark jet (defined as the mistag rate). Type-3 events are required to have at least two AK4 jets tagged at the ``medium'' working point of the algorithm with a \PQb\ tagging efficiency of 63\% and a 1\% mistag rate. In all cases, \PQb\ tagging algorithms are applied before removal of any double-counted leptons within AK4 jets.

The kinematic quantities of the top quark pair are reconstructed from their constituent decay products using a maximum-likelihood fit~\cite{Erdmann:2013rxa,James:1975dr}, which varies the momenta of the decay products within their resolutions and iterates over all possible assignments of jets to determine the most likely value of the unknown longitudinal momentum of the neutrino and the best possible assignment of jets that conforms with the decay hypothesis. In type-1 events, the top quark that decays to all-jets is assumed to be the \PQt-tagged AK8 jet, and the lepton side of the decay is reconstructed from all combinations of the lepton, \ptmiss, and any \PQb-tagged AK4 jets that result in a lepton+jets top quark candidate (lepton + \ptmiss\ + AK4 jet) separated from the \PQt-tagged AK8 jet by $\DR(\PQt_{\text{jets}},\PQt_{\text{lep}}) > 2$. Type-2 and -3 events ignore AK8 jets entirely in their reconstruction, and instead their possible jet configurations comprise all assignments of the four or five highest \pt\ AK4 jets to either the lepton or all-jets side of the decay; events containing five AK4 jets consider additional configurations in which any of the non-{\PQb}-tagged jets may be disregarded as external to the top quark pair decay.

For each event, the hypothesis with the smallest value of $-2\ln{L}$ (where $L$ is the likelihood) is chosen to represent the complete top quark pair decay, and the lepton and all-jets top quark four-vectors are reconstructed as the vector sums of their rescaled particle four-momenta. This kinematic fitting procedure is highly effective, returning the correct hypotheses in 98, 80, and 73\% of type-1, -2, and -3 lepton+jets \ttbar MC events, respectively, for which an MC-based particle-matched hypothesis exists.

After reconstruction, type-1 and type-2 events are further required to have top quark masses $m_{\PQt,\text{lep}}^{\text{reco}} < 210\GeV$ for decays containing one lepton, and $-2\ln{L} < -15$. The inversion of either of these criteria defines a control region used to constrain the cross section of the {\PW}+jets background process as described in Section~\ref{sec:background}. From applying the selection criteria to simulated events, we expect about $6210\pm80$ ($3110\pm60$), $39100\pm200$ ($2490\pm50$), and $188500\pm400$ ($134500\pm400$) \ttbar lepton+jets events in the type-1, -2, and -3 {\PGm}({\Pe})+jets signal regions, respectively. The relatively poor efficiency for type-2 electrons as compared with that for type-1 electrons results from the more stringent \ptmiss\ and lepton and jet \pt\ criteria applied to electron events in those regions and the lower \mtt\ of events in the type-2 signal region.

\begin{figure}[hbt]
  \centering
    \includegraphics[width=0.49\linewidth]{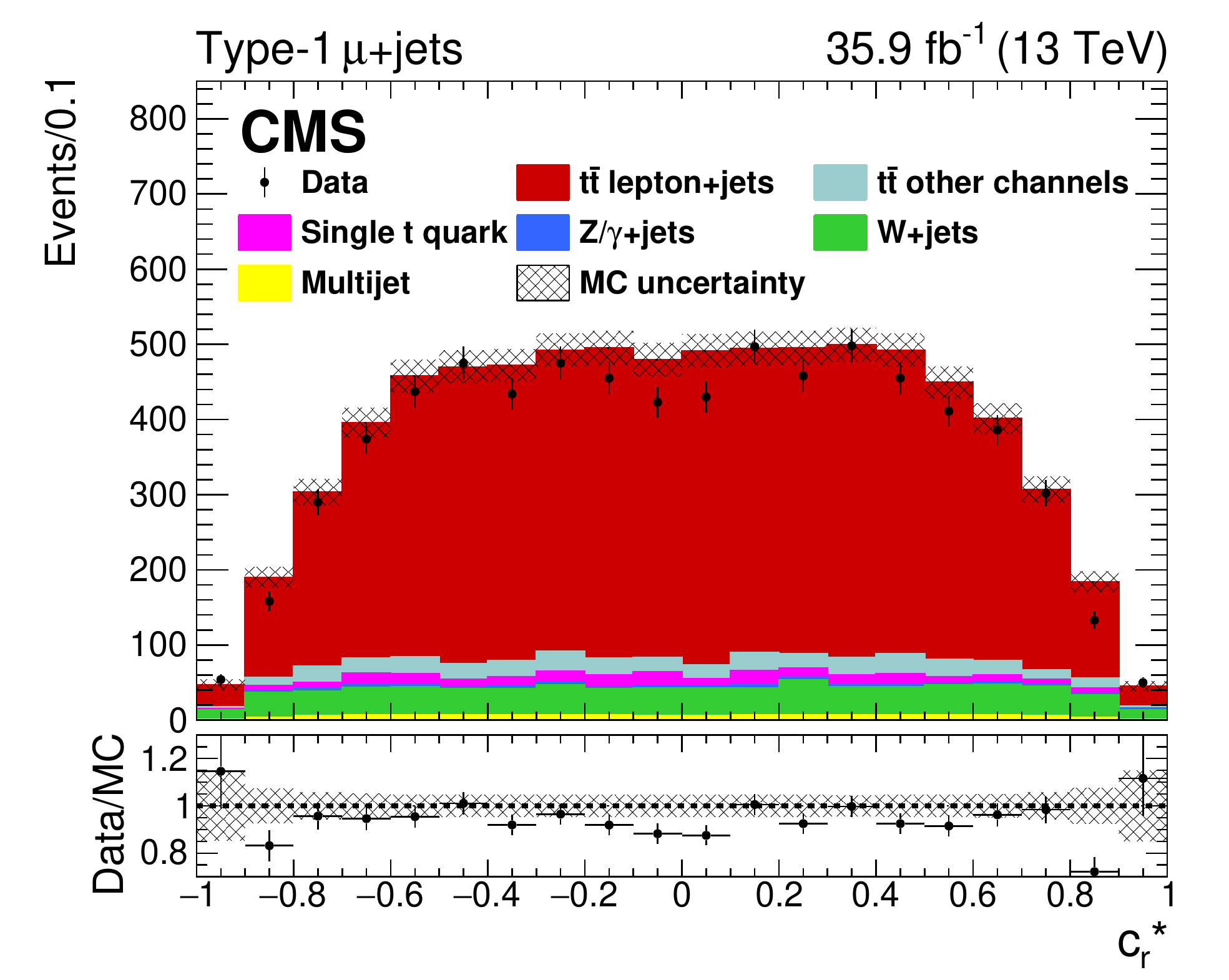}
    \includegraphics[width=0.49\linewidth]{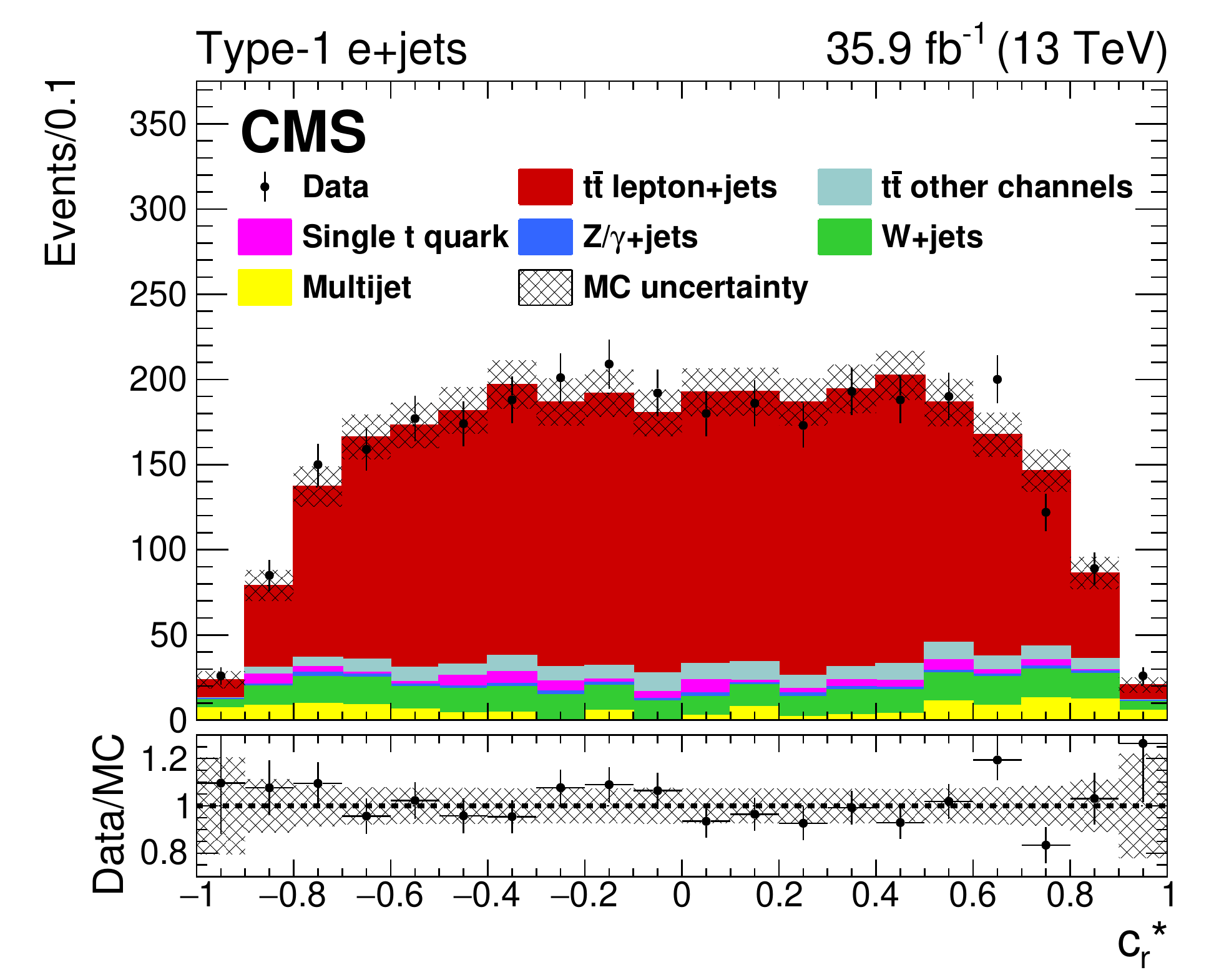}
    \includegraphics[width=0.49\linewidth]{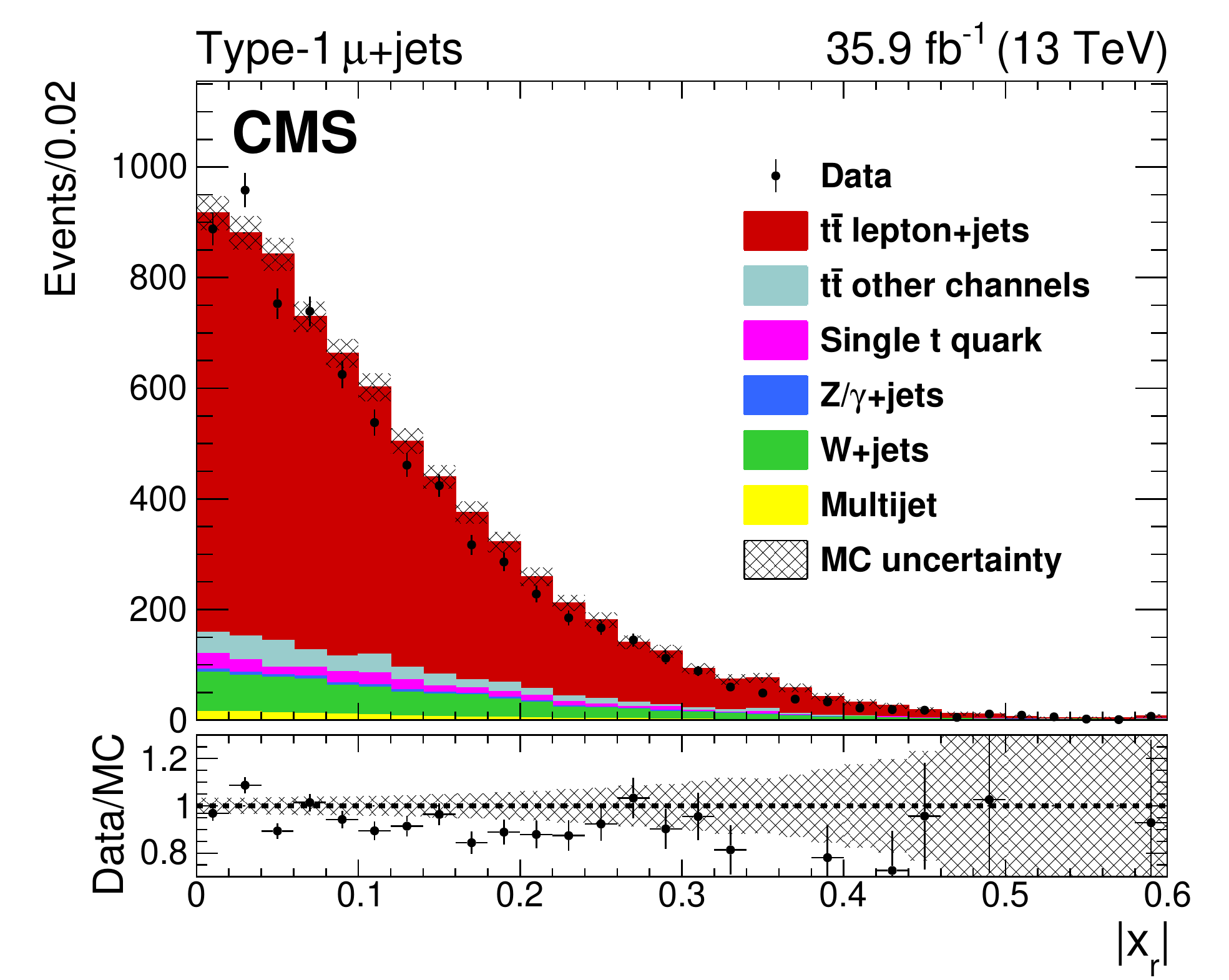}
    \includegraphics[width=0.49\linewidth]{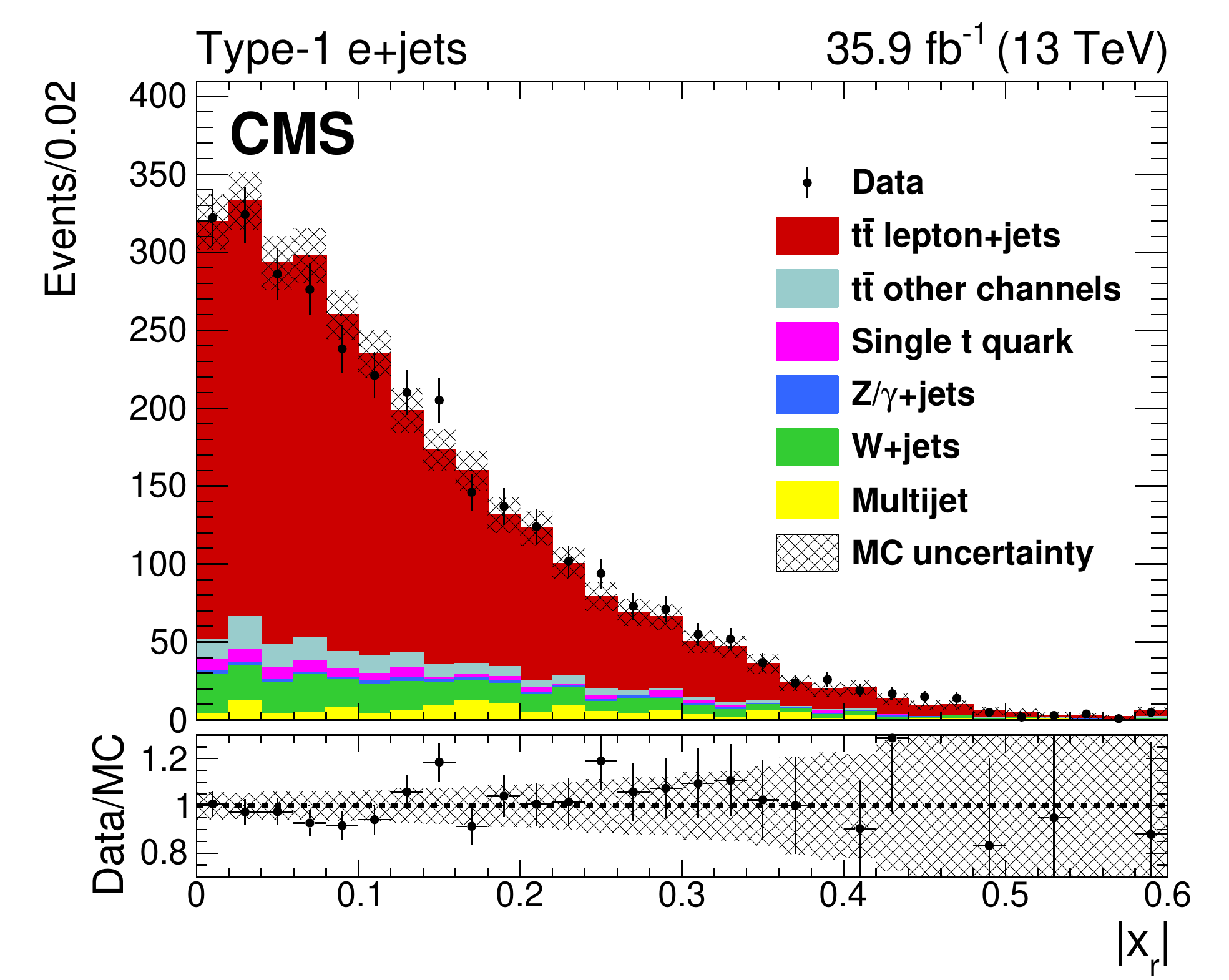}
    \includegraphics[width=0.49\linewidth]{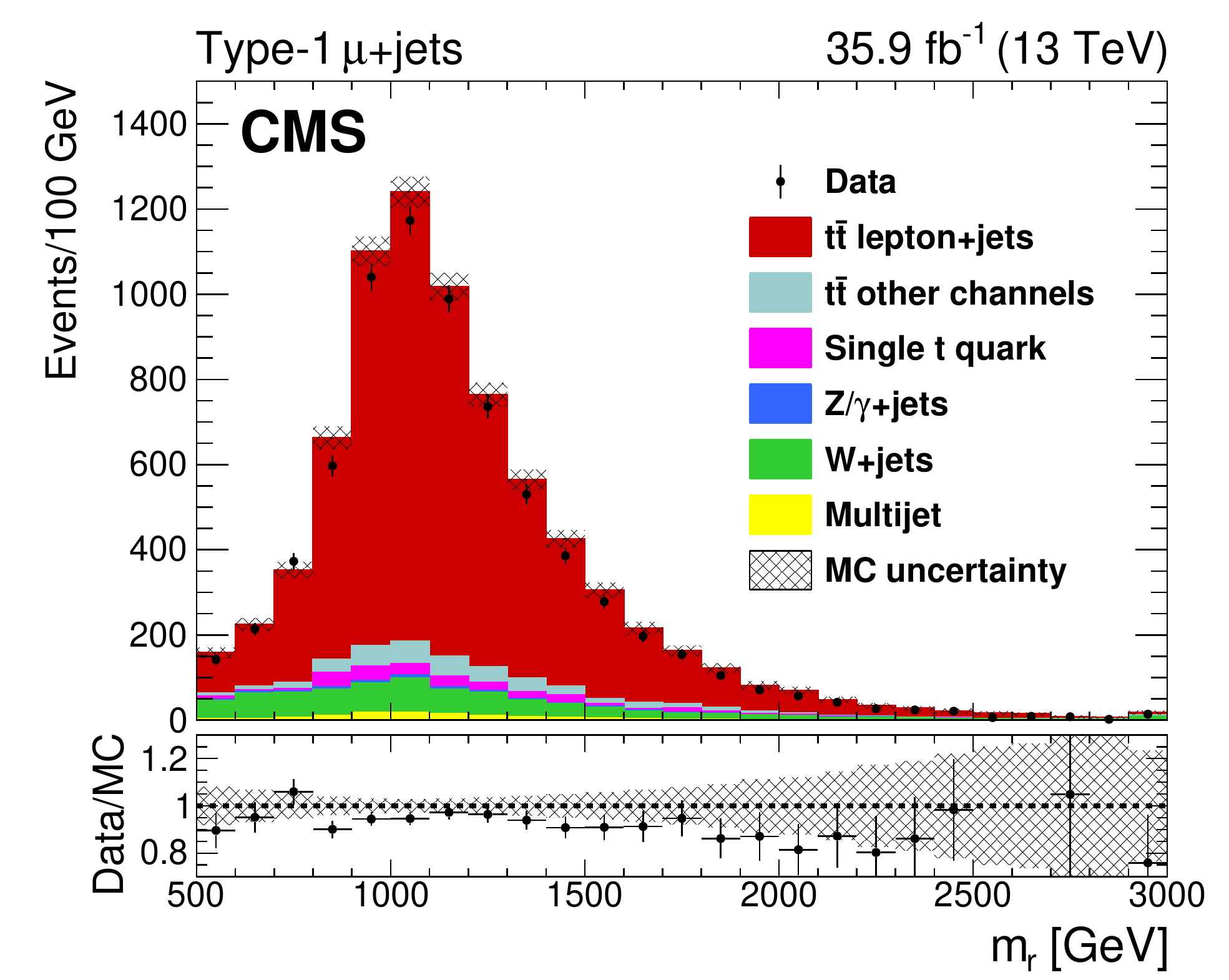}
    \includegraphics[width=0.49\linewidth]{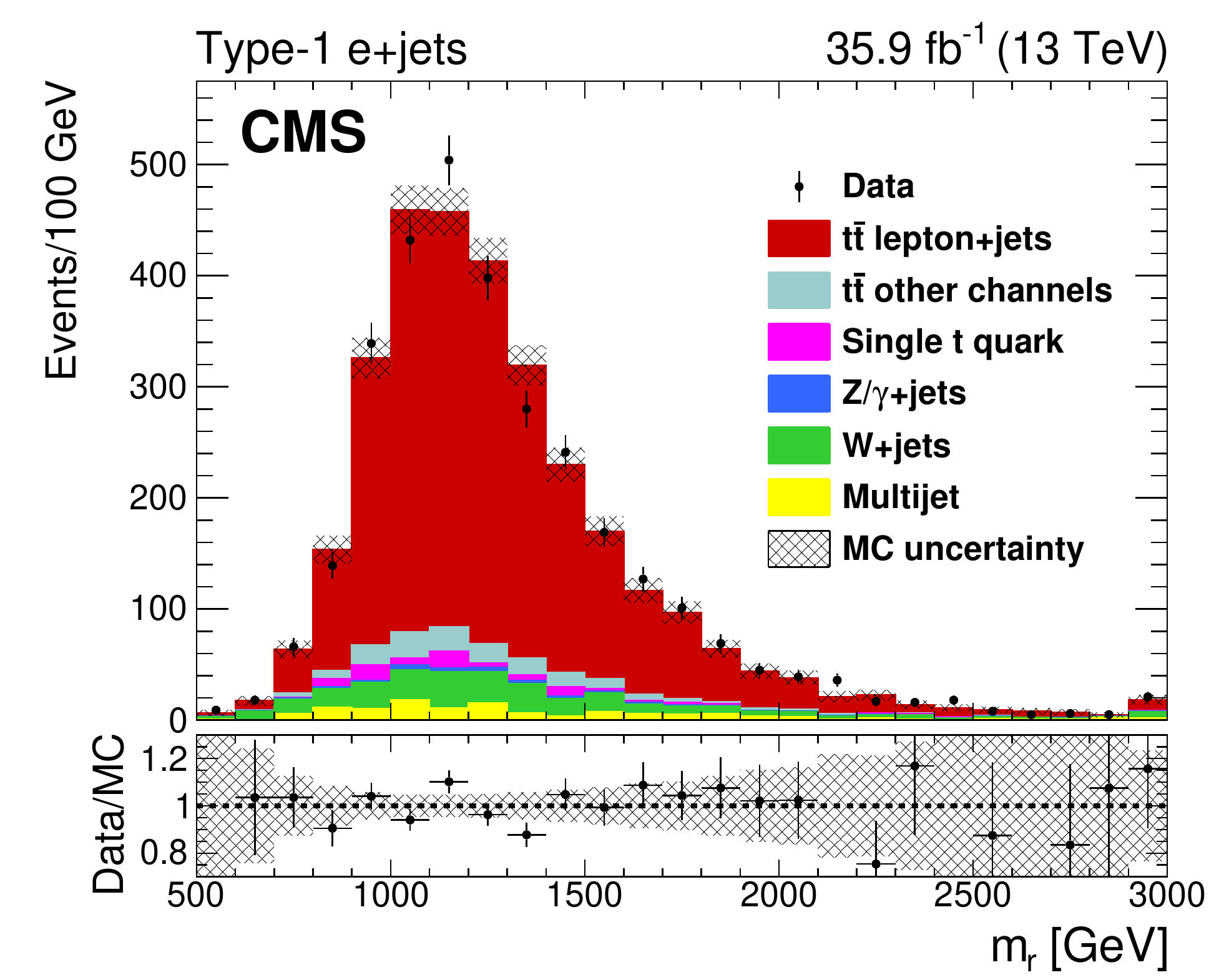}
  \caption{ Reconstructed \csubr\ (upper), $\abs{\xr}$ (middle), and \mr\ (lower) for events passing type-1 {\PGm}+jets (left column) and {\Pe}+jets (right column) selection criteria. The uncertainty pictured in the hatched bands represents just the statistical contributions. The multijet background is estimated from data, as discussed in Section~\ref{sec:background}. The lower panels in each figure show the ratio of data to MC expectation in each bin, and the last bins of the $\abs{\xr}$ and \mr\ plots include overflow.}
  \label{fig:control_plots_t1}
\end{figure}

\begin{figure}[hbt]
  \centering
    \includegraphics[width=0.49\linewidth]{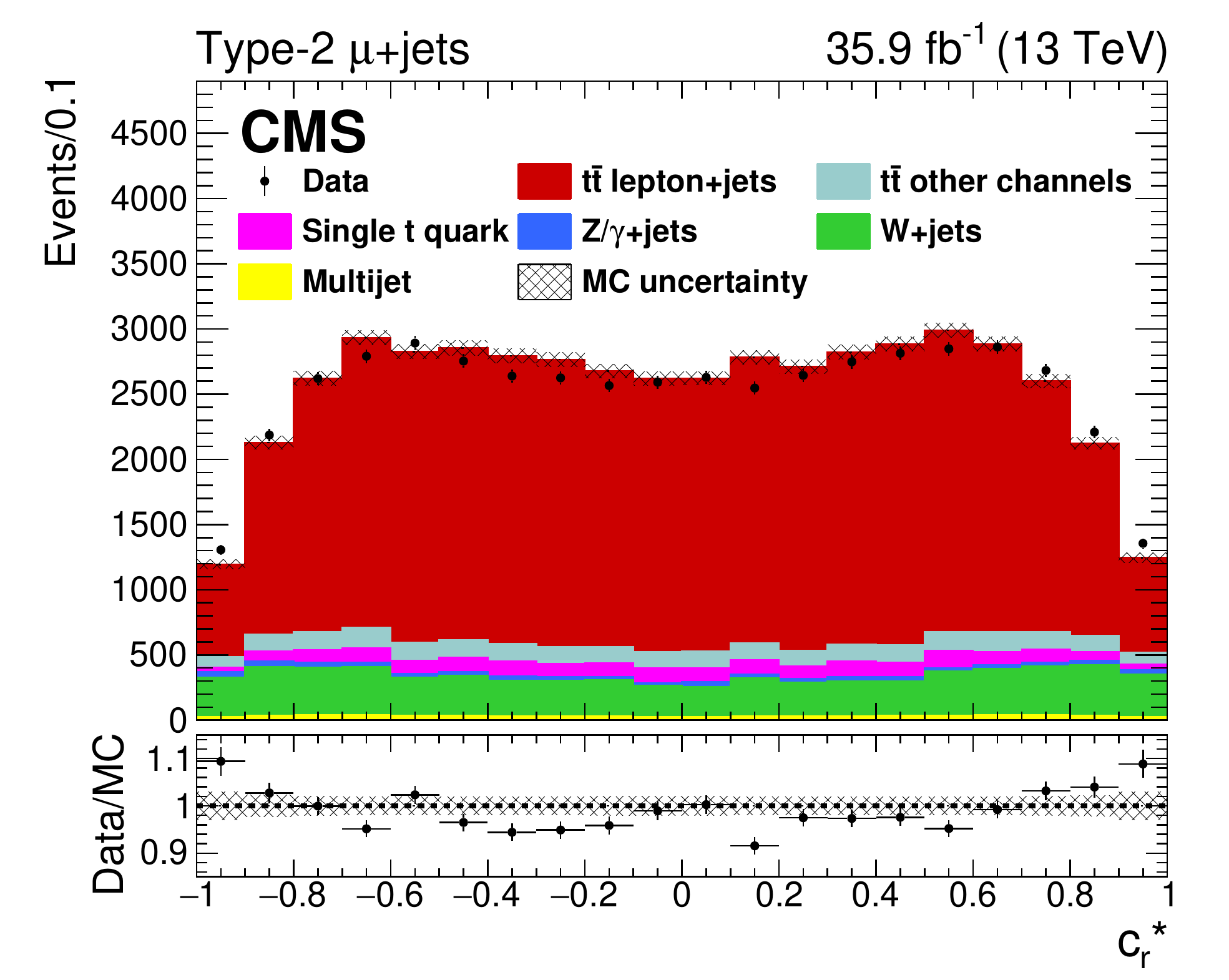}
    \includegraphics[width=0.49\linewidth]{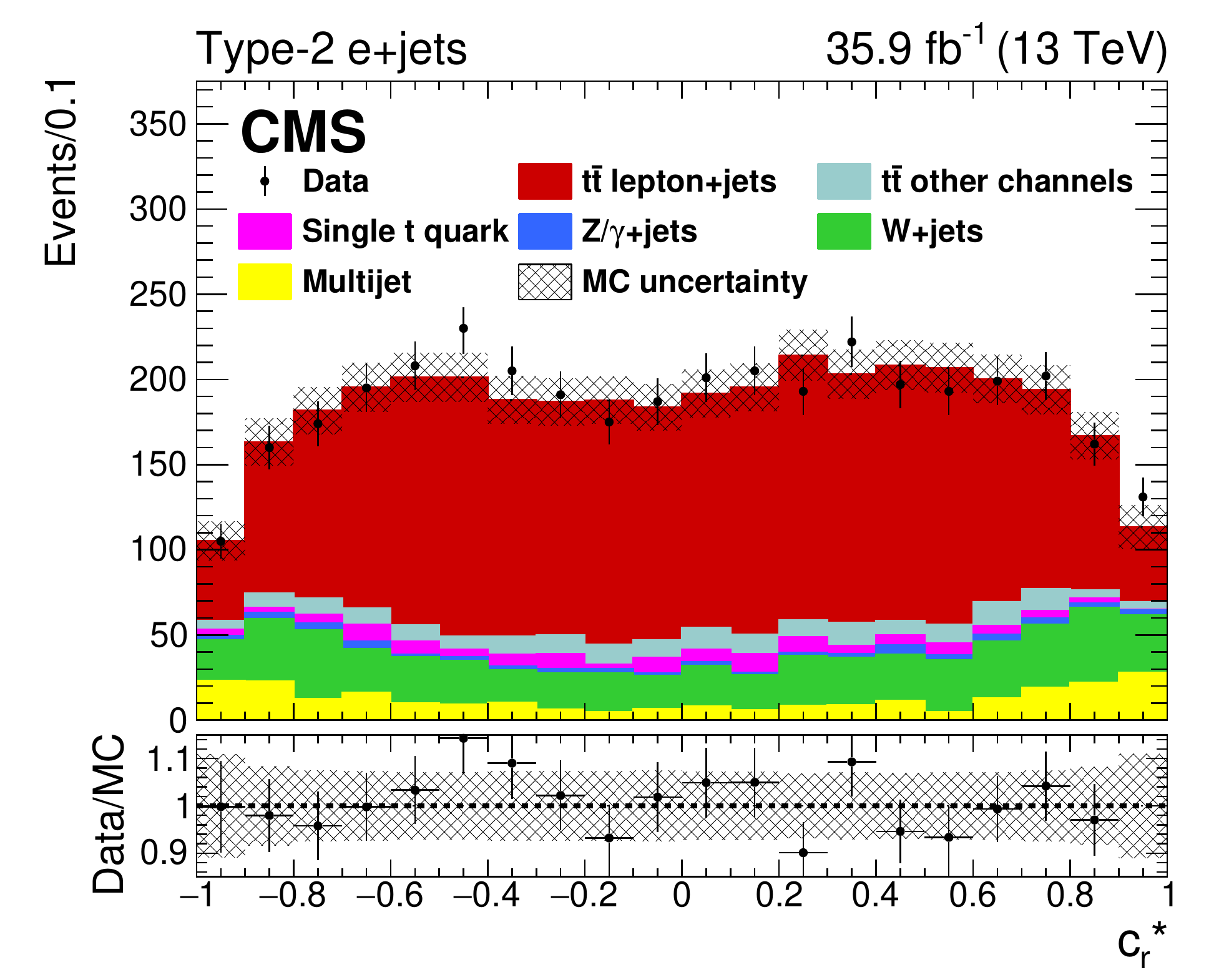}
    \includegraphics[width=0.49\linewidth]{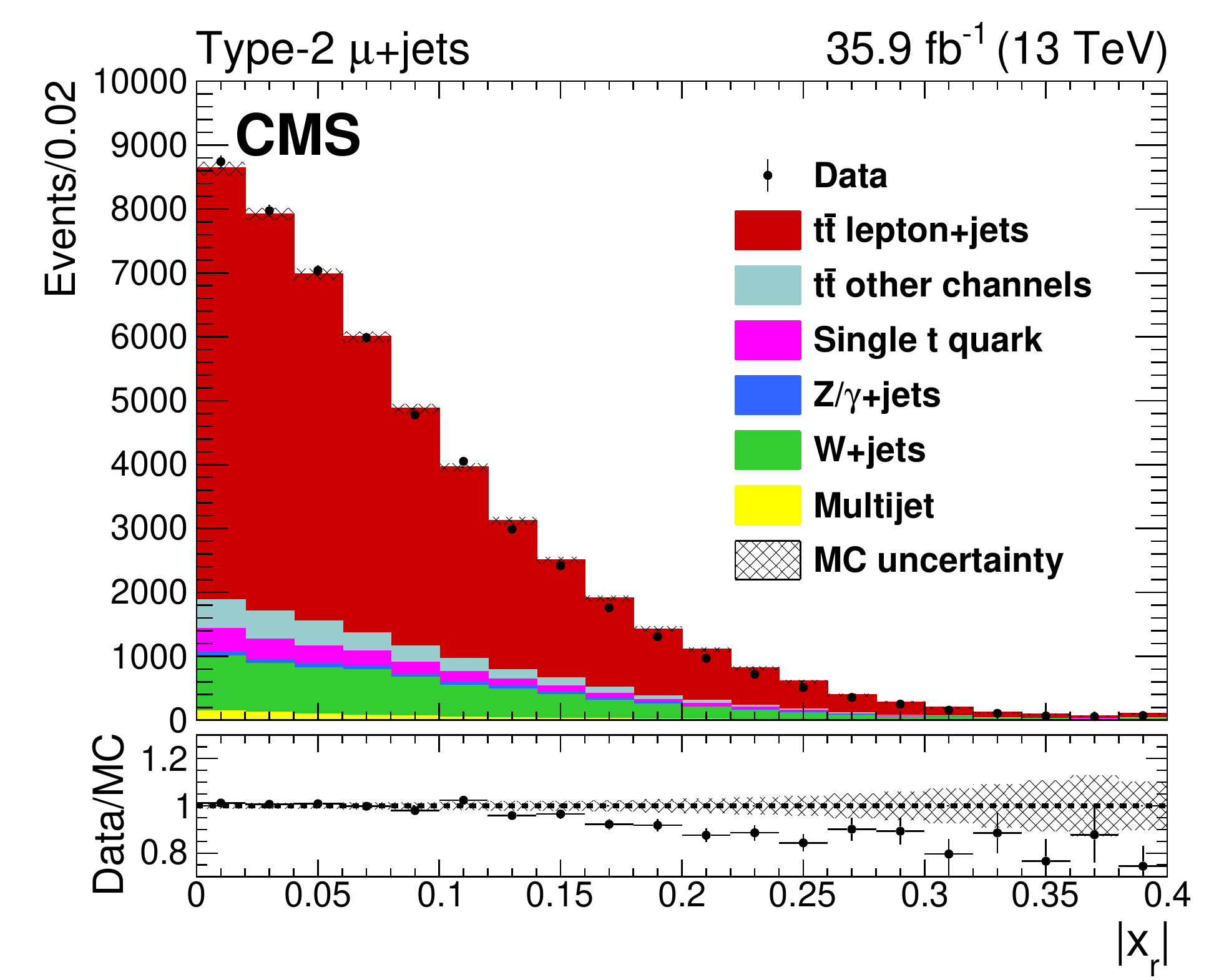}
    \includegraphics[width=0.49\linewidth]{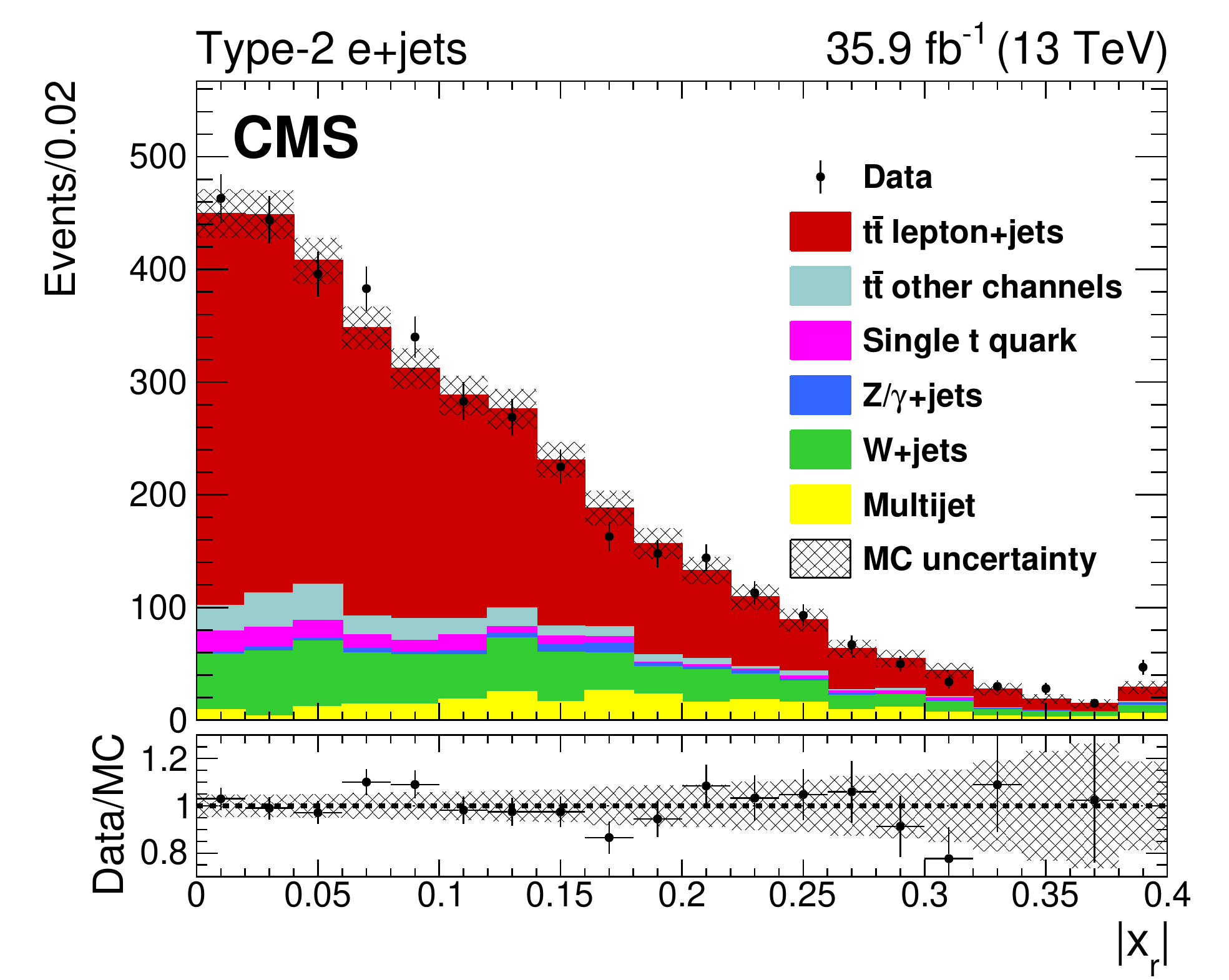}
    \includegraphics[width=0.49\linewidth]{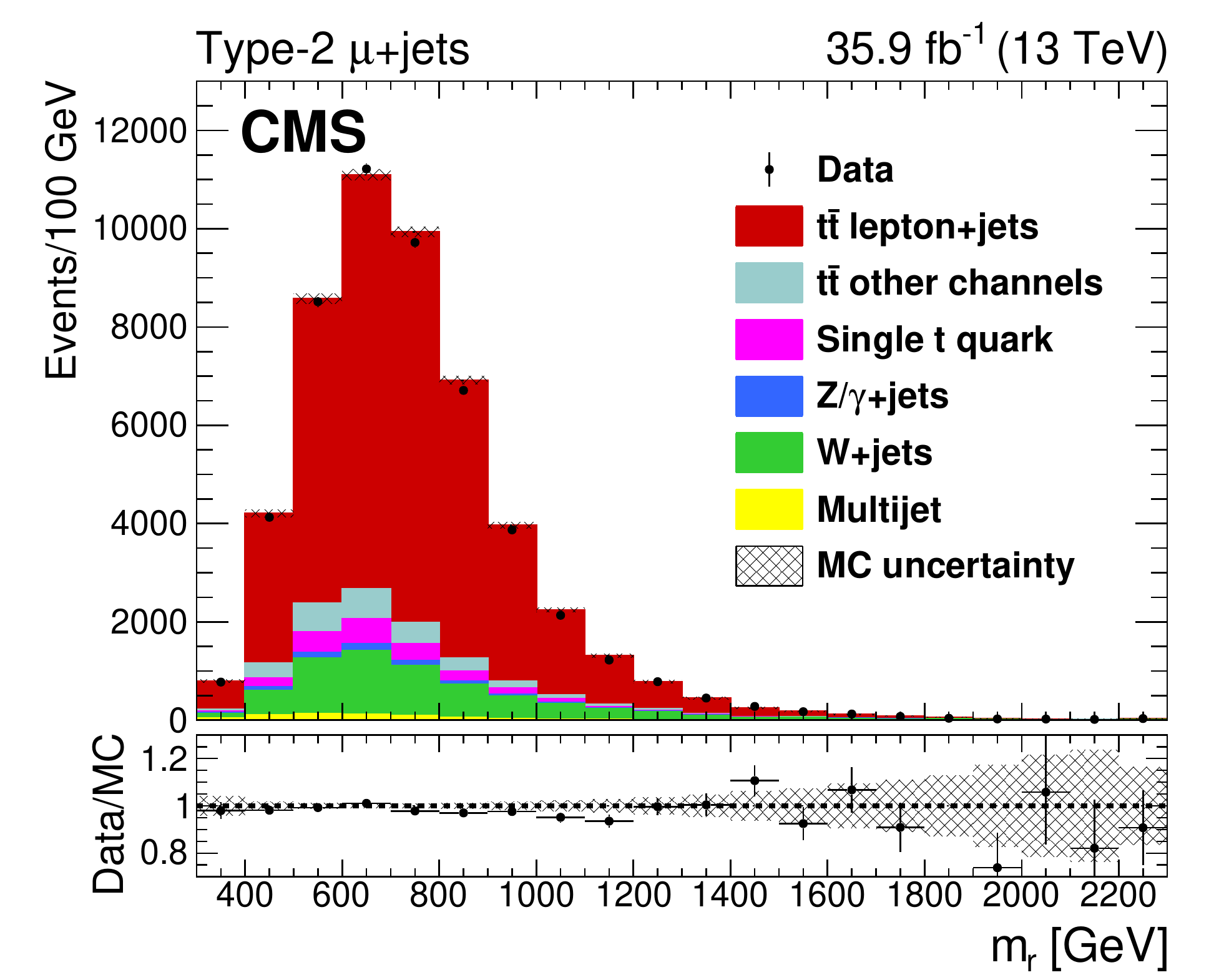}
    \includegraphics[width=0.49\linewidth]{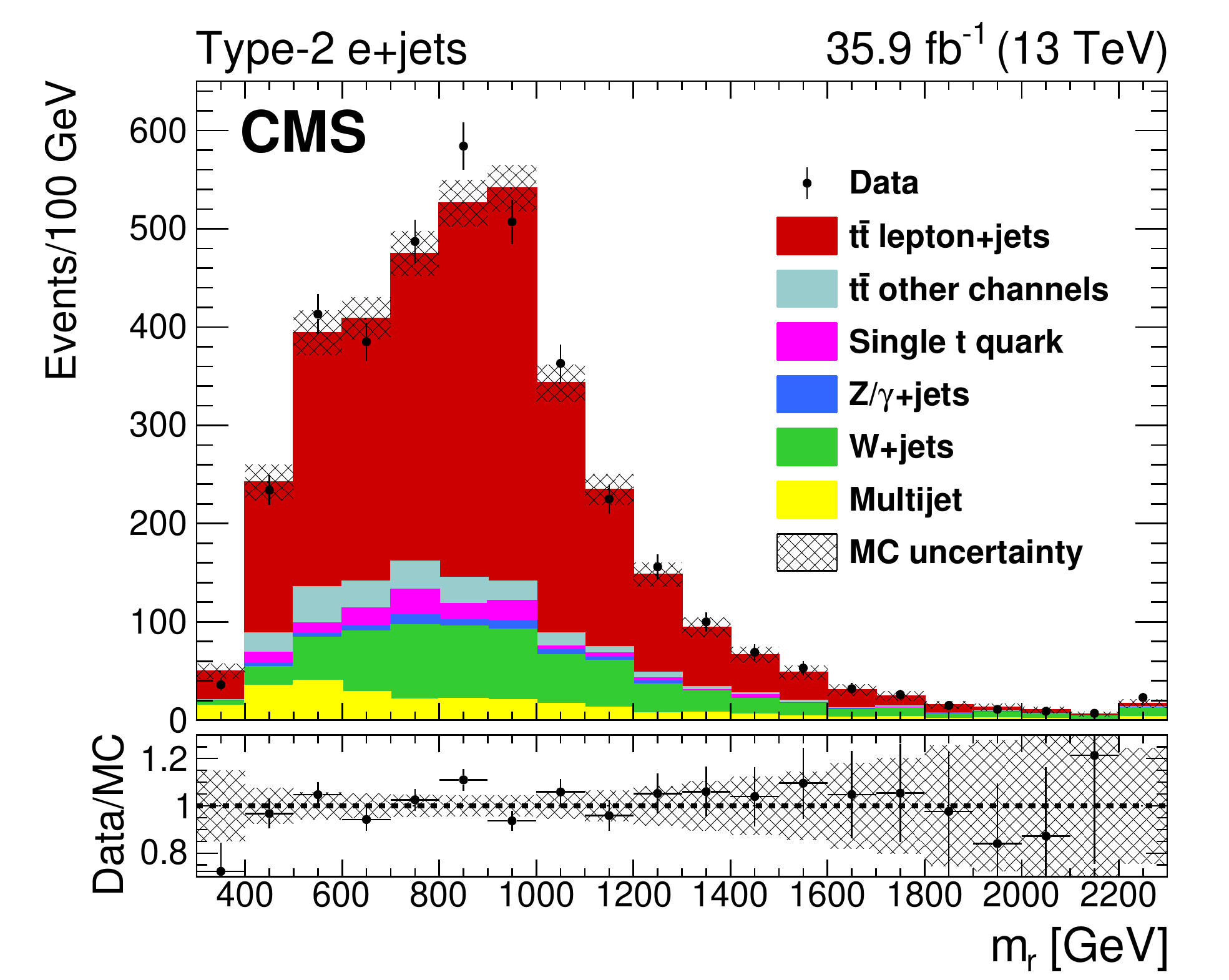}
  \caption{Data/MC comparison of reconstructed \csubr\ (upper), $\abs{\xr}$ (middle), and \mr\ (lower) for events passing type-2 {\PGm}+jets (left column) and {\Pe}+jets (right column) selection criteria. The MC signal and background show their nominal predictions, with the MC uncertainty pictured in the hatched bands representing statistical uncertainties. The contribution from multijet background is estimated from data, as discussed in Section~\ref{sec:background}. The lower panels of each figure show the ratio of the observed data to MC expectation in each bin, and the last bins of the $\abs{\xr}$ and \mr\ plots include overflow.}
  \label{fig:control_plots_t2}
\end{figure}

\begin{figure}[hbt]
  \centering
    \includegraphics[width=0.49\linewidth]{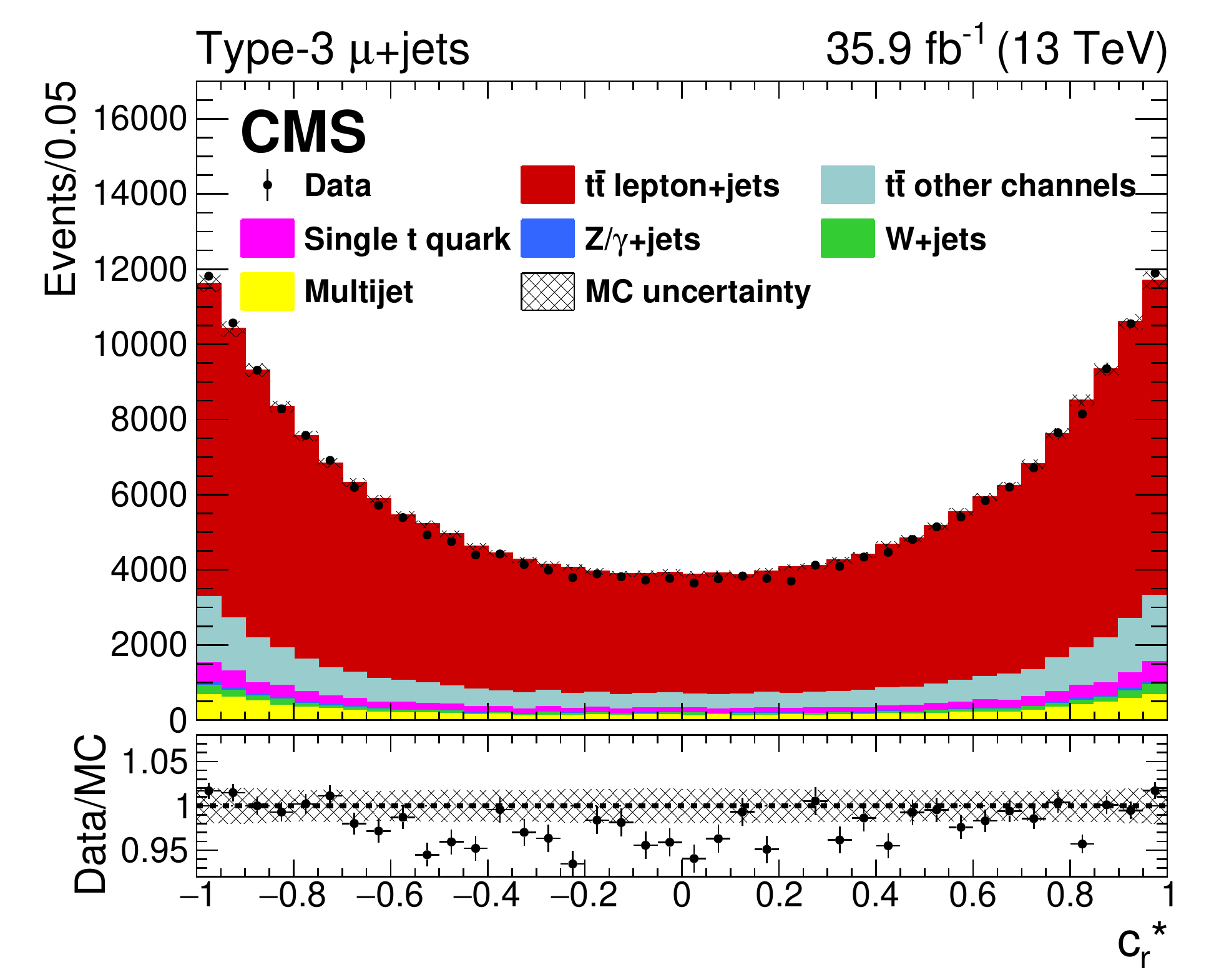}
    \includegraphics[width=0.49\linewidth]{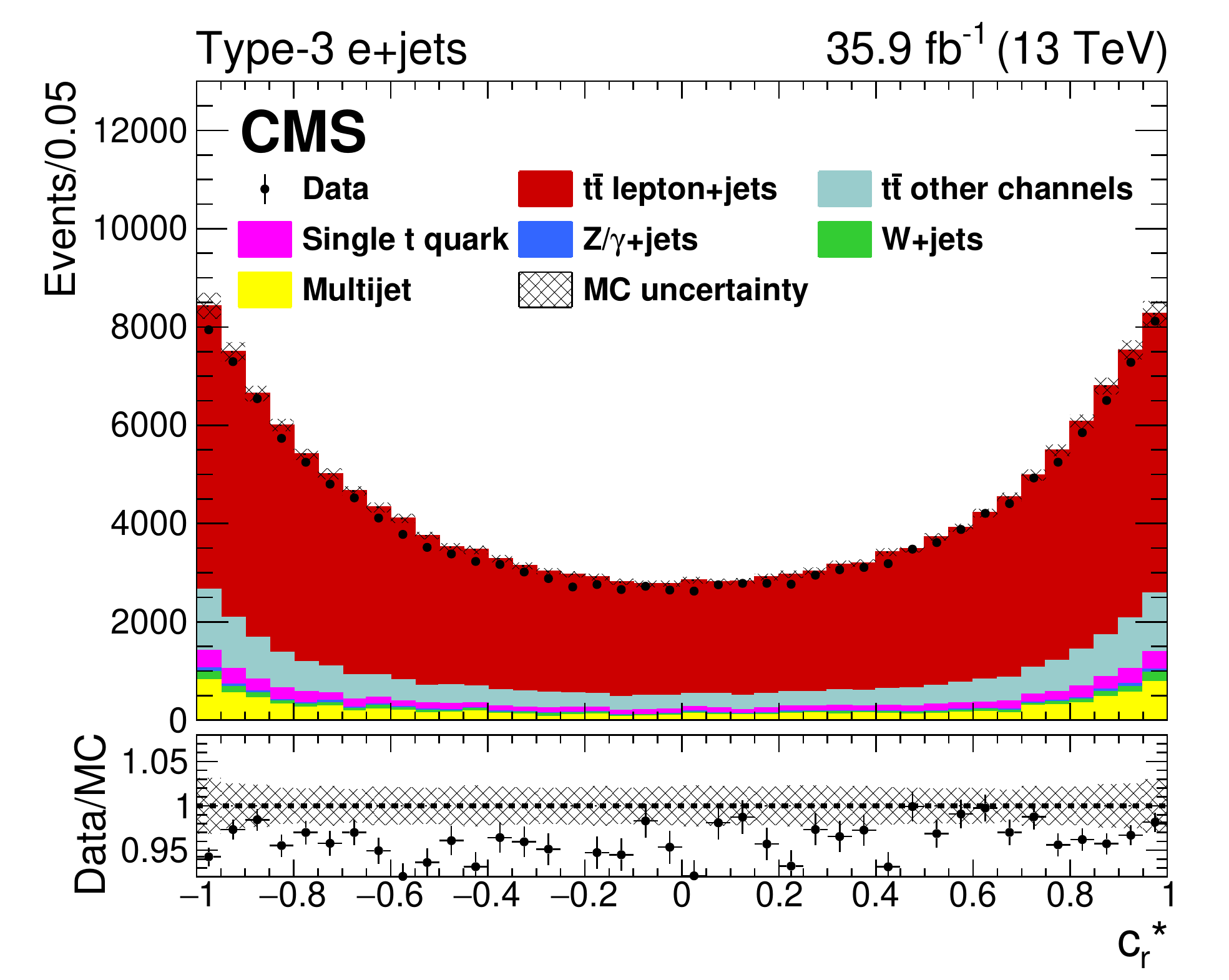}
    \includegraphics[width=0.49\linewidth]{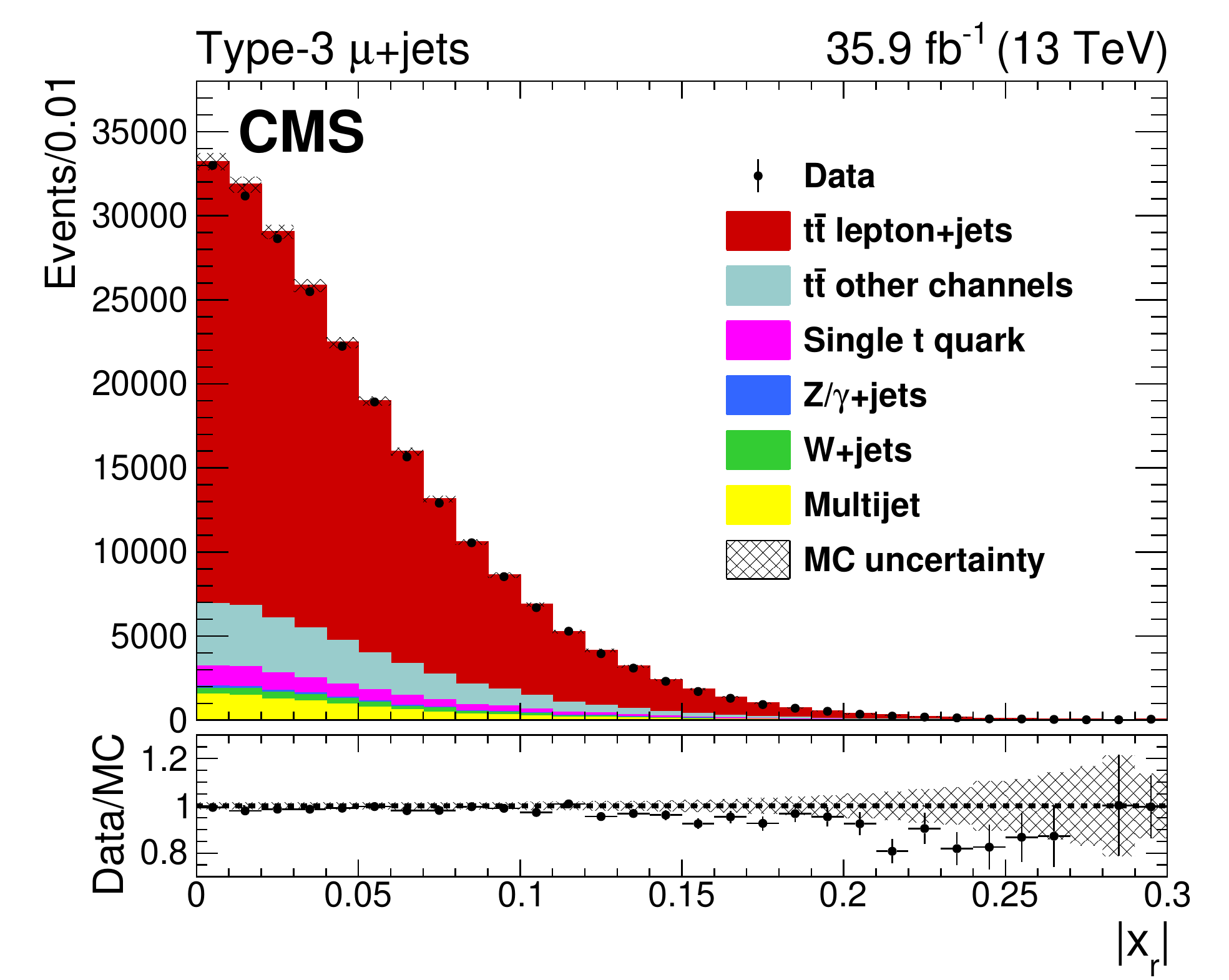}
    \includegraphics[width=0.49\linewidth]{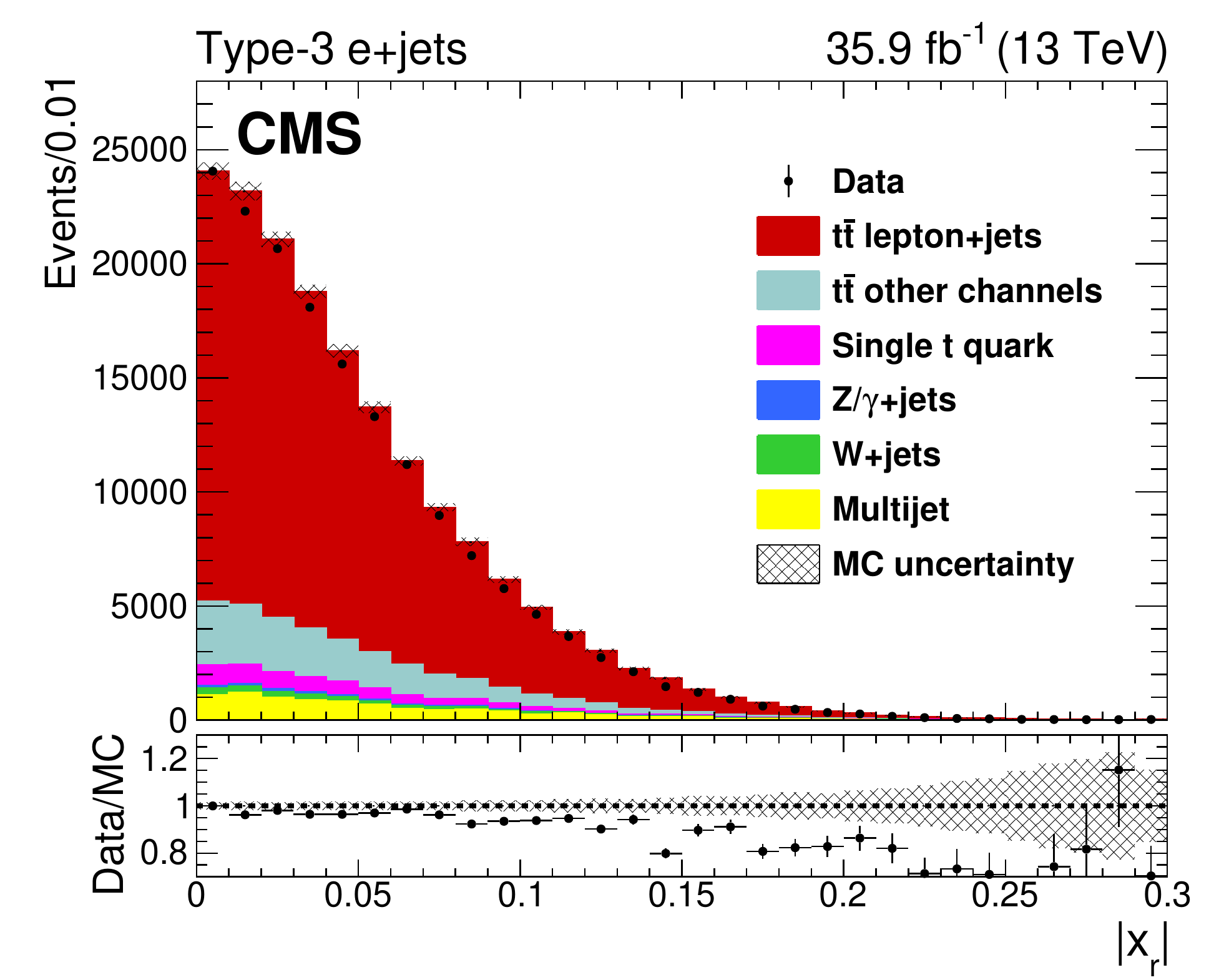}
    \includegraphics[width=0.49\linewidth]{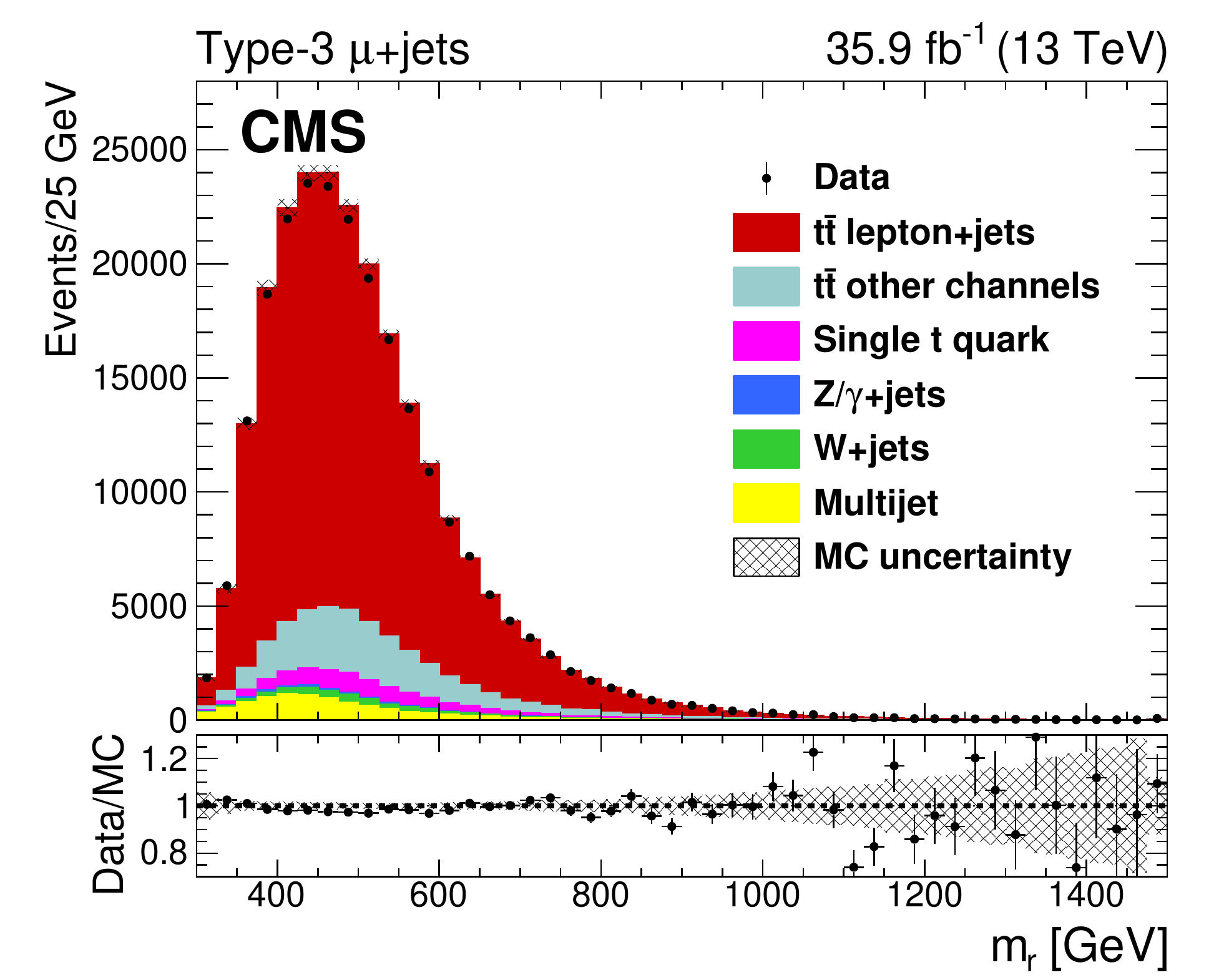}
    \includegraphics[width=0.49\linewidth]{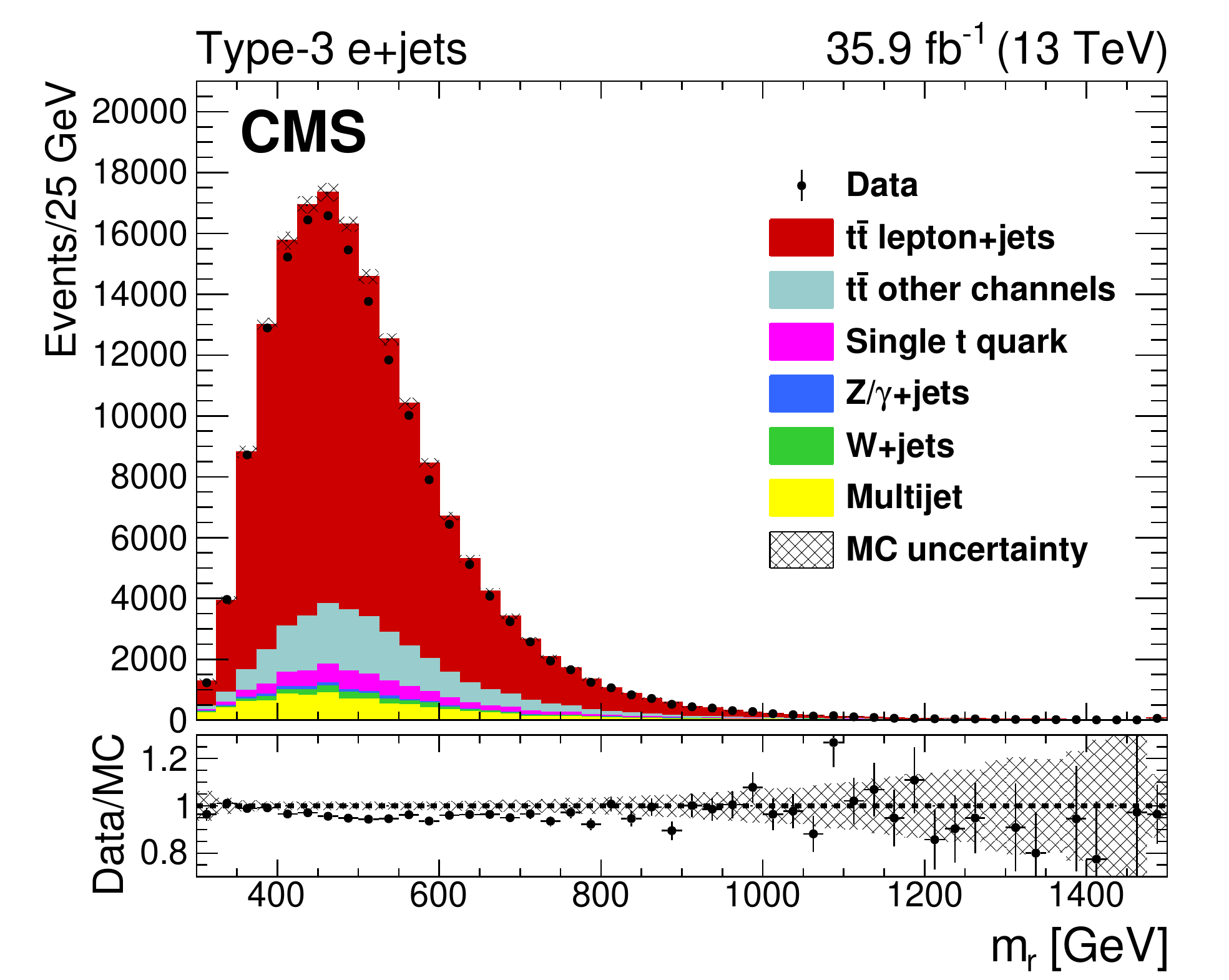}
  \caption{Data/MC comparison of reconstructed \csubr\ (upper), $\abs{\xr}$ (middle), and \mr\ (lower) for events passing type-3 {\PGm}+jets (left column) and {\Pe}+jets (right column) selection criteria. The MC signal and background show their nominal predictions, with the MC uncertainty pictured in the hatched bands representing statistical uncertainties. The contribution from multijet background is estimated from data, as discussed in Section~\ref{sec:background}. The lower panels of each figure show the ratio of the observed data to MC expectation in each bin, and the last bins of the $\abs{\xr}$ and \mr\ plots include overflow.}
  \label{fig:control_plots_t3}
\end{figure}

Figures~\ref{fig:control_plots_t1}--\ref{fig:control_plots_t3} show comparisons of the reconstructed variables, called \csubr, \xr, and \mr\ to distinguish them from the corresponding parton-level quantities defined in Section~\ref{sec:analysis_strategy}, for selected MC events and data for type-1, -2, and -3 regions, including the QCD multijet background estimated from data, as discussed in Section~\ref{sec:background}. The data-based method used to determine the multijet background contributions tends to overestimate the observed data, particularly in the type-3 electron sample; also, the disagreements visible in the \xr\ distributions are due to the choice of the PDFs.  Both of these systematic effects are corrected by the fit to data.

\clearpage

\section{Constructing templates and estimating background}
\label{sec:background}

To measure \aofb, the detector is assumed to have the same efficiency for reconstructing tracks of positively and negatively charged particles in the same configuration, implying a charge-parity symmetry where the acceptance for an event with a positively charged lepton and angles (\cstar, \csubr) is the same as that for a negatively charged lepton with angles ($-\cstar$, $-\csubr$). Counting data events reconstructed with positively- and negatively-charged leptons shows that this assumption is justified, with a maximum impact of about $10^{-5}$ on the \AFB\ quantity.

To exploit this symmetry, a fourth reconstructed quantity, the lepton charge $Q$, is used to describe each event.  The four-dimensional distribution function $f(\vec y)$ of the reconstructed variables $\vec y = (\mr, \csubr, \xr, Q)$ is determined from fully simulated and reconstructed events and from empirically determined reconstructed background,
\begin{linenomath}\begin{equation}
f(\vec y) = \sum_nR_{\text{bk},n}f_{\text{bk},n}(\vec y)+\left\lbrace \left(1-R_{\qqbar}\right) f_{\glu}(\vec y)+R_{\qqbar}\left[f_\text{qs}(\vec y)+\aofb f_\text{qa}(\vec y)\right]\right\rbrace ,
\label{eq:template_schemeone}
\end{equation}\end{linenomath}
where the $f_{\text{bk},n}(\vec y)$ represent normalized distribution functions for several backgrounds; $R_{\text{bk},n}$ are matching background-fraction scale factors; $R_{\qqbar}$ is the fraction of \ttbar events that are categorized as \qqbar\ initiated; $f_{\glu}(\vec y)$ is the normalized distribution function for \glu, \qg, {\Pq}{\Pq}, {\Paq}{\Paq}, and $\Pq_{i}\Paq_{j}$ (where $i$, $j$ label flavors and $i \neq j$) events; $f_\text{qs}(\vec y)$ is the symmetrized distribution for \qqbar\ events, and $f_\text{qa}(\vec y)$ is the antisymmetrized linearized distribution for \qqbar\ events.  The symmetrized function is created by increasing the event counts in the bins at (\csubr, $Q$) and ($-\csubr$, $-Q$) by 0.5 for each generated event. The antisymmetrized, linearized function in Eq.~(\ref{eq:qqonedef}) is constructed by adding half the weight
\begin{linenomath}\begin{equation}
w_\text{qa}(\mtt^2,\cstar) = \frac{2\left[2-\frac{2}{3}\beta^2+\alpha(1-\frac{1}{3}\beta^2)\right]\cstar}{2-\beta^2+\beta^2\cstarsq+\alpha\left(1-\beta^2\cstarsq\right)}
\label{eq:fqa_reweighting_factor}
\end{equation}\end{linenomath}
to the bin at (\csubr, $Q$) and half the weight $w_\text{qa}(\mtt^2, -\cstar)$ to the mirror bin at ($-\csubr$, $-Q$).  The values of the function $\alpha$ used in the weighting procedure are determined from fits to generator level data in 10 bins of $\beta$.  The resulting values of $\alpha$ are small, $\approx0.12$, except near the \ttbar\ threshold.

These templates are independent of any of the parameters to be determined.  This technique determines the average parton-level linearized asymmetry and accounts for resolution, dilution, migration, and acceptance effects, as long as they are modeled in the simulation.  The templates modeling well-simulated background contributions can likewise be populated using simulated events.

To measure \pard\ and \parmu, Eqs.~(\ref{eq:qqtwodef}) and~(\ref{eq:ggonedef}) are used to define a 4D distribution function in terms of eight parameter-independent template functions,
\begin{linenomath}\begin{equation}\begin{aligned}
f(\vec y) =&  \sum_nR_{\text{bk},n}f_{\text{bk},n}(\vec y)+\biggl\lbrace\frac{(1-R_{\qqbar})}{F_{\glu}(\parmu,\pard)}  \Bigl[ f_{\mathrm{g}0}(\vec y)+\parmu(1+\parmu)f_{\mathrm{g}1}(\vec y) \\
&+(\parmu^2+\pard^2)(1+\parmu)f_{\mathrm{g}2}(\vec y)+(\parmu^2+\pard^2)(1-5\parmu)f_{\mathrm{g}3}(\vec y)+(\parmu^2+\pard^2)^2f_{\mathrm{g}4}(\vec y)\Bigr]  \\
&+\frac{R_{\qqbar}}{F_{\qqbar}(\parmu,\pard)}\Bigl[ f_{\mathrm{q}0}(\vec y)+\left(2\parmu+\parmu^2-\pard^2\right)f_{\mathrm{q}1}(\vec y)+\left(\parmu^2+\pard^2\right))f_{\mathrm{q}2}(\vec y) \Bigr]\biggr\rbrace   ,
\label{eq:sixparone}
\end{aligned}\end{equation}\end{linenomath}
where the template functions $f_{\text{qi}}(\vec y)$ and $f_{\text{gi}}(\vec y)$ are constructed from simulated \ttbar events using weights, and the functions $F_{\qqbar}$ and $F_{\glu}$ maintain normalization of the \qqbar\ and \glu\ contributions independent of the parameters \pard\ and \parmu, as described in Appendix~\ref{app:reweighting_factors}.

The distribution functions in Eqs.~(\ref{eq:template_schemeone}) and~(\ref{eq:sixparone}) are fitted to data as templates in three-dimensional histograms in \mr, \xr, and \csubr\ for each reconstructed top quark pair, separated according to lepton charge $Q$. The histograms are binned differently in each channel, using variable bin widths in each dimension. Because the general analyzing power of any particular event is most dependent on its value of $\abs{\xr}$, a one-dimensional binning is first chosen in this quantity alone. A two dimensional binning in \csubr\ and \mr\ is then determined in each bin of $\abs{\xr}$. To maintain analyzing power for the \aofb\ parameter of interest, all areas of the phase space have a minimum of two bins depending on the sign of \csubr. Each template is stored as a list of two-dimensional histograms indexed by bin in $\abs{\xr}$, and is unrolled into a one dimensional histogram for fitting. Tables~\ref{tab:template_binning_t1_muons}--\ref{tab:template_binning_t3_electrons} in Appendix~\ref{app:template_binning} list the edges of the bins in $\abs{\xr}$ and their corresponding two dimensional \csubr\ and \mr\ binning schemes for each topology and lepton flavor.

The $f_\text{qs}(\vec y)$, $f_\text{qa}(\vec y)$, $f_{\glu}(\vec y)$, $f_{\text{qi}}(\vec y)$, and $f_{\text{gi}}(\vec y)$ templates are constructed by accumulating lepton+jets \ttbar MC events in each bin, reweighted to factorize out parameters of interest (\ie, through the weight in Eq.~(\ref{eq:fqa_reweighting_factor}) used for $f_\text{qa}(\vec y)$, and the weights in Appendix~\ref{app:reweighting_factors} used for $f_{\text{qi}}(\vec y)$ and $f_{\text{gi}}(\vec y)$). These templates, along with $f_{\text{bk},n}(\vec y)$ describing background contributions, are summed in a linear combination to estimate the total observed cross section.

The backgrounds contributing to this analysis are from dilepton and all-jet \ttbar, single top quark, Drell--Yan {\PZ}/{\PGg}+jets and {\PW}+jets, and multijet processes. Background contributions other than multijet events are estimated using MC predictions after applying corrections to account for differences in identification and selection efficiency between simulation and data. The top quark and {\PZ}/{\PGg}+jets backgrounds are considered together as one set of background templates $f_{\text{bk},1}(\vec y)$, which are expected to contribute approximately $670\pm30$ ($270\pm20$), $5300\pm70$ ($400\pm20$), and $36800\pm200$ ($26900\pm200$) events to the type-1, -2, and -3 {\PGm}({\Pe})+jets signal regions, respectively.

The {\PW}+jets background process is expected from simulation to contribute approximately $690\pm30$ ($260\pm20$), $6170\pm80$ ($570\pm20$), and $273200\pm500$ ($1940\pm40$) events to the type-1, -2, and -3 signal regions, respectively. The contribution to the background from {\PW}+jets processes is modeled using a dedicated set of background templates, $f_{\text{bk},2}(\vec y)$, because its production cross section is less precisely known from theory. For type-1 and -2 events, which expect larger relative {\PW}+jets contributions, the amount of {\PW}+jets background is constrained by performing a simultaneous fit to data in orthogonal control regions enriched in {\PW}+jets events. These control regions are populated by events that are otherwise selected as described in Section~\ref{sec:selection}, but which fail either of the requirements on $m_{\PQt,\text{lep}}^{\text{reco}}$ or $-2\ln{L}$, \ie, $m_{\PQt,\text{lep}}^{\text{reco}} \geq 210\GeV$ or $-2\ln{L} \geq -15$. This selection is chosen by comparing simulated semileptonic \ttbar and {\PW}+jets events; {\PW}+jets events consistently exhibit larger values of the kinematic fit and have leptonic top quark masses that are further away from the true top quark mass than semileptonic \ttbar events.

The QCD multijet background contribution represents a small fraction of the total hadronic final state cross section and may not be well described in simulation. This background is estimated from data using a matrix method similar to that of Ref.~\cite{Chatrchyan:2012cx}. The distribution in $\vec y$ of the multijet background is estimated in each channel using orthogonal selection sidebands, which contain events whose lepton candidates fail isolation requirements. The distributions are initially normalized by applying transfer factors, calculated as the ratios of the numbers of events that satisfy lepton isolation requirements to those that do not, measured in sidebands containing events that fail \ptmiss\ requirements. The expected residual contamination from background processes other than multijets is estimated from simulation and subtracted from the control samples. The transfer factors are approximately 0.004 (0.46), 0.003 (0.35), and 0.37 (1.44), resulting in nominal expectations of approximately $140\pm10$ ($170\pm10$), $1460\pm40$ ($330\pm20$), and $10200\pm100$ ($8610\pm90$) events, in the type-1, -2, and -3 signal region {\PGm}({\Pe})+jets channels, respectively. The final normalization of the multijet background templates $f_{\text{bk},3}(\vec y)$ is determined by the fit to data. Comparisons between the multijet background estimated from data and simulation show good agreement to within statistical uncertainties.

\section{Systematic uncertainties}
\label{sec:systematics}

Systematic uncertainties in the normalization and shapes of the template distributions arise from a variety of sources and are accounted for through ``nuisance'' parameters that include their statistical limitations, and their constraints by the prior distributions described below. These sources and the methods of their accounting are listed below and summarized in Table~\ref{tab:nuisance_listing}.

\noindent\textit{Jet energy corrections:} As described in Section~\ref{sec:detector}, jet energies and their resolutions are corrected in the simulation to agree with observations in data. The jet energy corrections are achieved through the application of a scale factor that depends on the jet \pt, $\eta$, area $A$, and the event pileup $n_{\text{PV}}$.  The jet energy resolution is corrected by the application of a Gaussian smearing that depends on the jet $\abs{\eta}$. The jet energy scale and resolution factors are independently shifted up and down by one standard deviation ($\sigma$) as they are applied to AK4 and AK8 jets simultaneously, including the propagation of the corrections to the observed \ptmiss.  The up and down variations are then used to construct sets of up and down templates for use in the fitting procedure. The resulting templates are smoothed to reduce bias from statistical noise before fitting.

\noindent\textit{Pileup:} Uncertainties in the procedure used for reweighting the pileup distribution in the simulation are included by varying the inelastic {\Pp}{\Pp} cross section ${\pm}4.6\%$~\cite{Sirunyan:2018nqx}.

\noindent\textit{Trigger and lepton identification and isolation efficiencies:} Trigger efficiencies in data and simulation are determined using multiple independent methods, each resulting in a lepton \pt- and $\eta$-dependent scale factor applied to simulated events. The different algorithms for determining lepton identification and isolation provide their own sets of scale factors dependent on pileup, lepton \pt, and lepton $\eta$. The average values of the scale factors are ${\approx}0.94$ and 0.96 for boosted and resolved {\PGm}+jets events, respectively, and ${\approx}0.98$ for boosted and resolved {\Pe}+jets events. The precision of the boosted {\Pe}+jets trigger efficiency measurement is limited by the number of events passing selection criteria. These scale factors are independently shifted up and down by one $\sigma$ to provide sets of templates for interpolation.

\noindent\textit{\PQb\ tagging efficiency:} Scale factors dependent on jet flavor, \pt, and $\eta$ are applied to simulated events to correct for differences in \PQb tagging efficiency and mistag rates between data and simulation. Their uncertainties are propagated to templates independently for each jet flavor and working point of the \PQb\ tagging algorithm.

\noindent\textit{Top tagging efficiency:} Three data/MC efficiency scale factors with associated systematic uncertainties are applied to all simulated events selected with \PQt-tagged jets to correct for the differences in tagging efficiency between data and simulation. The three factors applied depend on the jet \pt\ according to whether the particular jet is fully, partially, or not merged, as determined by an MC matching procedure.

\noindent\textit{Top quark \pt\ reweighting:} Recent NNLO QCD + electroweak calculations of top quark pair production~\cite{Czakon:2017wor} describe how NNLO effects impact the top quark and antiquark \pt\ spectra in ways that NLO generators cannot reproduce. Scale factors dependent on NLO-generated top quark and antiquark \pt\ are calculated and applied to bring the generated \pt\ distributions into agreement with those predictions, and the uncertainties in the calculated scale factors are considered as systematic uncertainties~\cite{Sirunyan:2019wka}.

\noindent\textit{PDFs and strong coupling:} The systematic uncertainty from the choice of the NNPDF3.0 PDFs used in generating \ttbar MC events is included using the ${\pm}1\sigma$ deviations observed in 100 per-event weights representing the uncertainties in the PDFs~\cite{Ball:2014uwa}. Changes in \alpS\ are included by recalculating the generated event weight with $\alpS=0.118 \pm 0.0015$~\cite{Butterworth:2015oua}. This uncertainty is combined in quadrature with the overall PDF uncertainty to provide one set of up/down templates describing the simultaneous variation of PDF and \alpS\ \cite{Rojo:2015acz,Accardi:2016ndt}.

\noindent\textit{Renormalization and factorization scales:} Modeling uncertainties on the renormalization (\muR) and factorization (\muF) scales used in the matrix element generation process are included by reweighting simulated events to match alternate scenarios with \muR\ and \muF\ shifted up and down by a factor of two, both independently and simultaneously, resulting in three nuisance parameters describing \muR, \muF, and combined \muR\ and \muF\ scales~\cite{Cacciari:2003fi,Catani:2003zt}.

\noindent\textit{Parton shower radiation, matching scales, and the underlying event:} Uncertainties in initial-state radiation (ISR) and final-state radiation (FSR), and matrix element to parton shower (ME-PS) matching, as well as uncertainties from the choice of CUETP8M2T4 tune in the \ttbar simulation, are included using templates constructed from independent MC samples generated with \PYTHIA\ parameters shifted by their uncertainties~\cite{CMS-PAS-TOP-16-021}. The resulting templates are smoothed to reduce bias from statistical noise before fitting.

\noindent\textit{Modeling of color reconnection:} Systematic uncertainties from the choice of modeling color reconnection are included using a single set of up/down changes in templates constructed as an envelope of the average fractional shifts observed in each bin of templates constructed from independent MC samples corresponding to three different color reconnection hypotheses~\cite{Sjostrand:2014zea,Argyropoulos:2014zoa,Christiansen:2015yqa}. These templates are also smoothed before fitting.

\noindent\textit{\PQb\ quark fragmentation and \PB\ hadron semileptonic branching fraction:} Uncertainties in \PQb\ quark fragmentation and the \PB\ hadron semileptonic branching fractions are included using per-event scale factors that depend on the generator-level transfer function $x_{\PQb} = \pt(\text{\PB})/\pt(\PQb-\text{jet})$, where $\pt(\PB)$ represents the transverse momentum of the \PB\ hadron, for fragmentation uncertainties~\cite{Landsman:2003zz}, and on the ratios of measured to simulated branching fractions for branching fraction uncertainties~\cite{PDG2018}.

\noindent\textit{Integrated luminosity:} The total integrated luminosity has an uncertainty of ${\pm}2.5\%$~\cite{CMS-PAS-LUM-17-001} that corresponds to a single nuisance parameter represented by a log-normal prior correlated across all analysis channels.

\noindent\textit{Process yields:} A ${\pm}1\%$ uncertainty in the fraction of lepton+jets top quark-pair production caused by quark-antiquark annihilation is included as a nuisance parameter $R_{\qqbar}$ affecting the distributions in \ttbar templates. A ${\pm}10\%$ uncertainty on the cross section of {\PW}+jets is included as a nuisance parameter $R_{\PW+\text{jets}}$, affecting the normalization of all $f_{\text{bk},2}(\vec y)$ templates. The scales of these uncertainties are chosen to characterize variations between different simulations. Their optimal values as determined by the fitting procedure are smaller, reflecting the constraints imposed by the observed data.

\noindent\textit{Transfer factors for multijet background estimated from data:} The estimated statistical uncertainties in the transfer factors for each multijet background estimate are modeled by several independent variation nuisances $R_{\text{QCD}}^{t/C/R}$, one for each estimate that is made. The scales of these uncertainties are approximately 30 and 20\% for type-1 {\PGm}+jets events with positively and negatively charged muons, respectively, and approximately 10\% for type-1 {\Pe}+jets events of both electron charges. For type-2 and -3 events of any lepton flavor and charge the uncertainties are on the order of a few percent.

\noindent\textit{Finite MC sample event count:} The statistical fluctuations in MC predictions are included through the ``Barlow--Beeston light" method~\cite{Conway:2011in}, which adds a Poisson uncertainty reflecting the number of events accumulated in each template bin.

\begin{table}[ht!]
\topcaption{ List of nuisance parameters considered in fits to data. The ``N" stands for ``normalization,'' and ``S'' for ``shape" of the distribution in the ``Type" column. The ``Size" column lists the absolute value of the associated fractional shifts averaged over all affected template bins. The quantities $R_{\text{QCD}}^{t/C/R}$ indicate that the QCD multijet yield uncertainties are independent in each topology, channel, and region.}
\centering
\begin{tabular}{ p{5.0 cm} c c c c }
Source & Uncertainty in & Type & Size & Affects \\
\hline
Jet energy scale                    & ${\pm}1\sigma(\pt,\eta,A)$                                   & N \& S & 7.6\% & All \\
Jet energy resolution               & ${\pm}1\sigma(\abs{\eta})$                                   & N \& S & 3.2\% & All \\
Pileup                              & ${\pm}1\sigma(n_{\text{PV}})$                                & N \& S & 2.9\% & All \\
Boosted {\PGm}+jets trigger eff.    & ${\pm}1\sigma(\pt,\eta)$                                     & N \& S & 0.4\% & Type-1/2 {\PGm}+jets \\
Resolved {\PGm}+jets trigger eff.   & ${\pm}1\sigma(\pt,\eta)$                                     & N \& S & 0.1\% & Type-3 {\PGm}+jets \\
Boosted {\Pe}+jets trigger eff.     & ${\pm}1\sigma(\pt,\abs{\eta})$                               & N \& S & 18.6\% & Type-1/2 {\Pe}+jets \\
Resolved {\Pe}+jets trigger eff.    & ${\pm}1\sigma(\pt,\eta)$                                     & N \& S & 2.5\% & Type-3 {\Pe}+jets \\
Muon ident. eff.                    & ${\pm}1\sigma(\pt,\abs{\eta},n_{\text{PV}})$                 & N \& S & 0.4\% & All {\PGm}+jets \\
Muon PF isolation eff.              & ${\pm}1\sigma(\pt,\abs{\eta},n_{\text{PV}})$                 & N \& S & 0.2\% & Type-3 {\PGm}+jets \\
Electron ident. eff.                & ${\pm}1\sigma(\pt,\abs{\eta})$                               & N \& S & 1.0\% & All {\Pe}+jets \\
\PQb tag eff., \PQb jets (loose)    & ${\pm}1\sigma(\pt,\eta)$                                     & N \& S & 2.5\% & Type-1/2 \\
\PQb tag eff., \PQc jets (loose)    & ${\pm}1\sigma(\pt,\eta)$                                     & N \& S & 1.2\% & Type-1/2 \\
\PQb tag eff., light jets (loose)   & ${\pm}1\sigma(\pt,\eta)$                                     & N \& S & 6.3\% & Type-1/2 \\
\PQb tag eff., \PQb jets (medium)   & ${\pm}1\sigma(\pt,\eta)$                                     & N \& S & 1.9\% & Type-3 \\
\PQb tag eff., \PQc jets (medium)   & ${\pm}1\sigma(\pt,\eta)$                                     & N \& S & 0.8\% & Type-3 \\
\PQb tag eff., light jets (medium)  & ${\pm}1\sigma(\pt,\eta)$                                     & N \& S & 1.2\% & Type-3 \\
\PQt tag eff. (merged)              & ${\pm}1\sigma(\pt)$                                          & N \& S & 1.6\% & Type-1 \\
\PQt tag eff. (semimerged)          & ${\pm}1\sigma(\pt)$                                          & N \& S & 2.2\% & Type-1 \\
\PQt tag eff. (not merged)          & ${\pm}1\sigma(\pt)$                                          & N \& S & 2.8\% & Type-1 \\
ISR scale                           & ${\pm}1\sigma$                                               & N \& S & 2.2\% & \ttbar \\
FSR scale                           & ${\pm}1\sigma$                                               & N \& S & 2.6\% & \ttbar \\
ME-PS matching ($h_{\text{damp}}$)  & ${\pm}1\sigma$                                               & N \& S & 2.5\% & \ttbar \\
CUETP8M2T4 tune                     & ${\pm}1\sigma$                                               & N \& S & 2.4\% & \ttbar \\
Color reconnection                  & ${\pm}1\sigma$                                               & S      & 2.8\% & \ttbar \\
\PQb fragmentation                  & ${\pm}1\sigma(x_{\PQb})$                                     & N \& S & 3.7\% & \ttbar \\
\PQb branching fraction             & ${\pm}1\sigma$                                               & N \& S & 1.0\% & \ttbar \\
Top quark \pt reweighting           & ${\pm}1\sigma(\pt^{\text{gen},\PQt},\pt^{\text{gen},\PAQt})$ & S      & 2.5\% & \ttbar \\
PDF/\alpS variation                 & NNPDF 3.0                                                    & S      & 1.5\% & \ttbar \\
Renormalization scale \muR          & $\frac{1}{2}\muR \to 2\muR$                                  & S      & 2.6\% & \ttbar \\
Factorization scale \muF            & $\frac{1}{2}\muF \to 2\muF$                                  & S      & 1.5\% & \ttbar \\
Combined \muR/\muF\ scale           & $\frac{1}{2} \to 2(\muR\text{ and }\muF)$                    & S      & 3.8\% & \ttbar MC \\
Integrated luminosity               & ${\pm}2.5\%$                                                 & N      & \NA   & All \\
$R_{\qqbar}$                        & ${\pm}1\%$                                                   & N \& S & \NA   & All $f_{\text{qp}*}/f_{\text{qm}*}$ \\
$R_{\PW+\text{jets}}$               & ${\pm}10\%$                                                  & N      & \NA   & All {\PW}+jets MC \\
$R_{\text{QCD}}^{t/C/R}$ (20 params total) & ${\pm}1\sigma\stat$                                   & N      & \NA   & Multijet \\
\hline
\end{tabular}
\label{tab:nuisance_listing}
\end{table}

A template-based, binned likelihood, combining the three-dimensional observable space in all channels and regions, is constructed to compare the data with expectation using a Poisson probability in each independent template bin as a function of the fit parameters. Systematic uncertainties are defined through nuisance parameters in the likelihood. Shape uncertainties are incorporated by interpolation between the nominal and shifted templates, constrained with a Gaussian prior. The interpolation is calculated with a sixth-order polynomial for shifts smaller than one $\sigma$, and with a linear function for shifts beyond one $\sigma$. Some shifted templates are smoothed before fitting by merging bins in \mr\ and \csubr\ and applying the average shifts. Normalization uncertainties are included using log-normal priors. The hierarchies of the most impactful systematic uncertainties are different for each parameter of interest, but one persistent observation is that the PDF/\alpS\ variation tends to be a large effect, correcting the \xF\ distribution to better agree with that observed in data.

These nuisance parameters and their corresponding functions are incorporated into three different likelihood fits to determine the three parameters of interest. A fit based on Eq.~(\ref{eq:template_schemeone}) is used to estimate \aofb\ and therefore assumes that $\pard = \parmu = 0$.  Fits based on Eq.~(\ref{eq:sixparone}) are used to separately extract \pard\ (with $\parmu = 0$) and \parmu\ (with $\pard = 0$) where we assume that $\AFB = 0.036$ (the value from the \ttbar MC). The resulting estimates are not sensitive to this assumption because \pard\ and \parmu\ affect only the {\cstar}-symmetric part of the cross section.

Final values of \aofb\ and \parmu\ are determined from a Neyman construction~\cite{10.2307/91337} in which 1000 pseudo-data sets are generated from the template models with input values of parameters of interest and then fitted. The median and ${\pm}1\sigma$ contours are plotted, and the value of the parameter of interest is the input value in the pseudo-experiments whose fitted median is the value returned by the fit to data, interpolating linearly between points if needed. The uncertainty intervals are constructed similarly from the ${\pm}1\sigma$ curves. When calculating the total intervals (statistical + systematic), the pseudo-data sets are generated using values of the nuisance parameters which are randomly sampled from their prior distributions. The intervals from which the statistical uncertainties are derived are instead obtained using pseudo-data sets in which nuisance parameters are fixed to their nominal values in both the toy generation and the fit. The final 95\% confidence limit on $\abs{\pard}$ is determined using a one-dimensional likelihood profile, because of the unresolved sign ambiguity of \pard\ in Eq.~(\ref{eq:sixparone}), and because of the small observed value of the \pard\ parameter. The upper limit at 95\% confidence level, computed using the asymptotic formulae for the distribution of the test statistic $\tilde{t}(\alpha)$ of Ref.~\cite{Cowan:2010js}, corresponds to a shift in the negative log likelihood of $-2\Delta \ln{L}=3.84$.

\section{Results}
\label{sec:results}

\begin{figure}[hbt]
  \centering
    \includegraphics[width=0.49\linewidth]{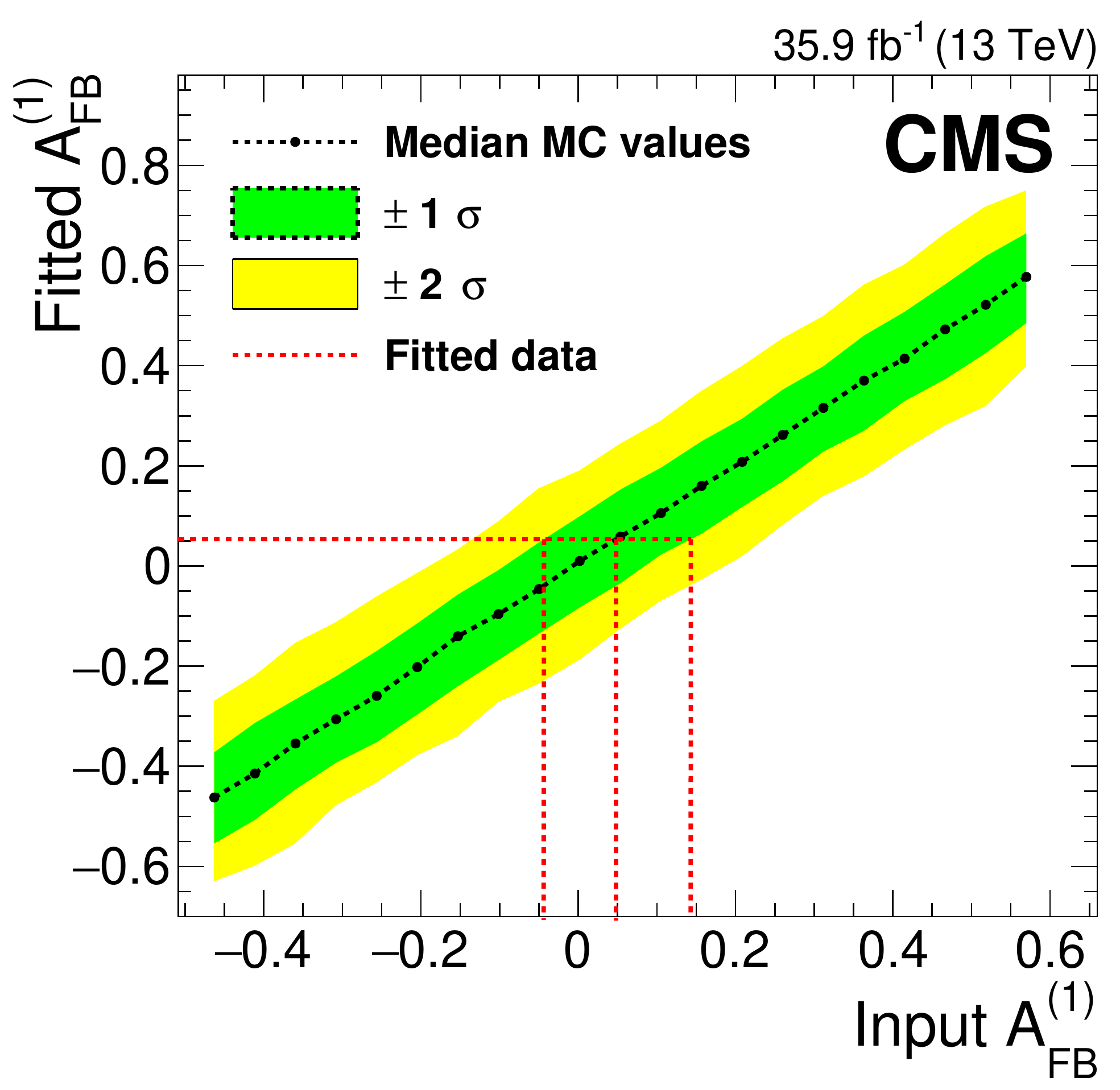}
    \includegraphics[width=0.49\linewidth]{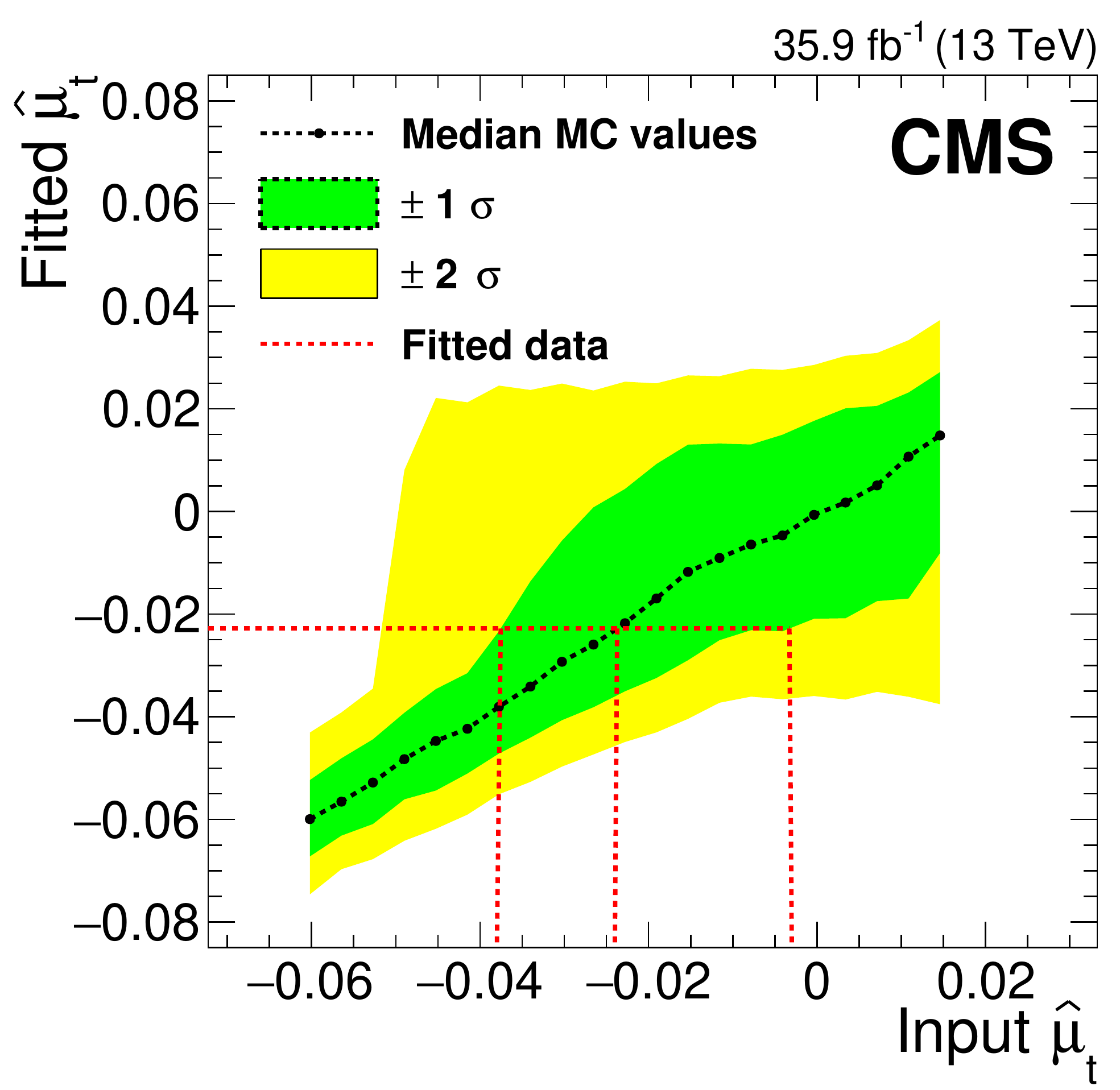}
  \caption{ Neyman constructions for the \aofb\ (left) and \parmu\ (right) parameters of interest in groups of 1000 pseudo-experiments generated with systematic uncertainty nuisance parameters allowed to vary. The horizontal dotted lines indicate the values of the parameters determined from the fits and the vertical dotted lines indicate where these values intersect with the central value and uncertainty contours from the pseudo-experiment groups.}
    \label{fig:neyman_plots}
\end{figure}

\begin{figure}[hbt]
  \centering
    \includegraphics[width=0.49\linewidth]{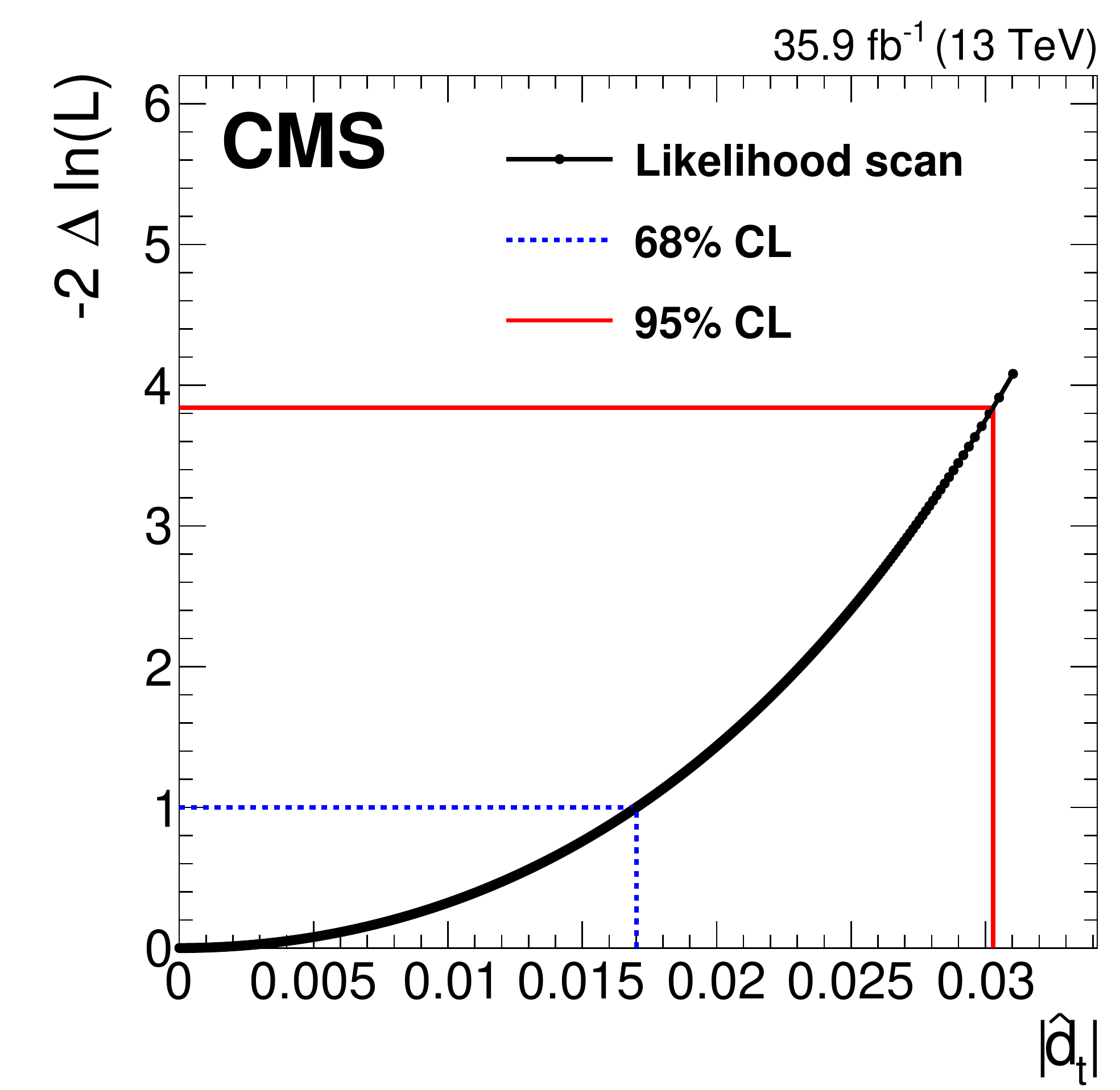}
  \caption{ One-dimensional likelihood profile of the \pard\ parameter, where the change in the minimum log likelihood from its global minimum ($-2\Delta \ln{L}$) is shown on the $y$-axis as a function of $\abs{\pard}$. The dashed and solid lines show the intersections at $-2\Delta \ln{L}=1.0$ and $3.84$, corresponding to the one-sided limits on $\abs{\pard}$ at the 68 and 95\% confidence limits, respectively.}
    \label{fig:d_likelihood_profile}
\end{figure}

\begin{figure}[hbt]
  \centering
    \includegraphics[width=0.49\linewidth]{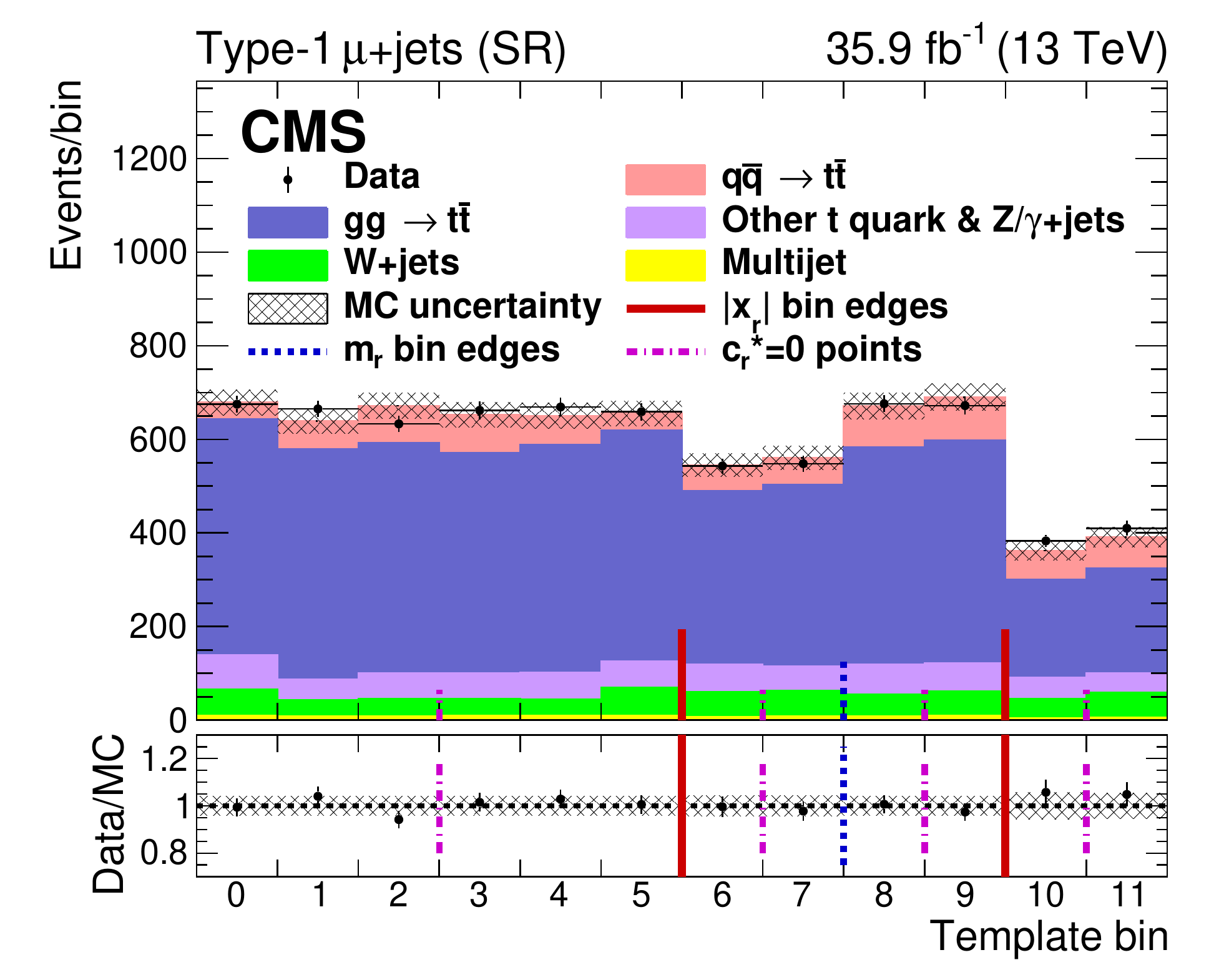}
    \includegraphics[width=0.49\linewidth]{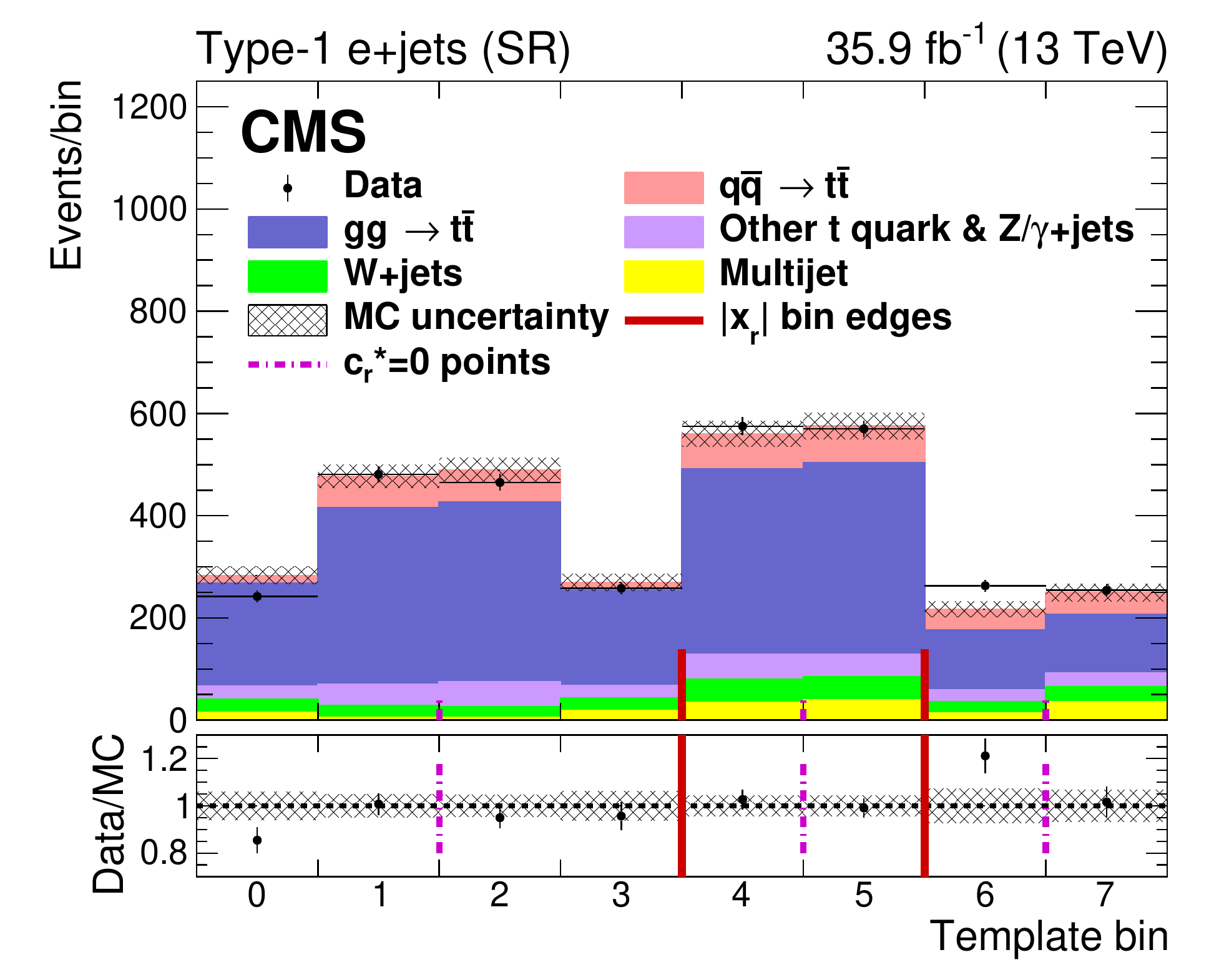}
    \includegraphics[width=0.49\linewidth]{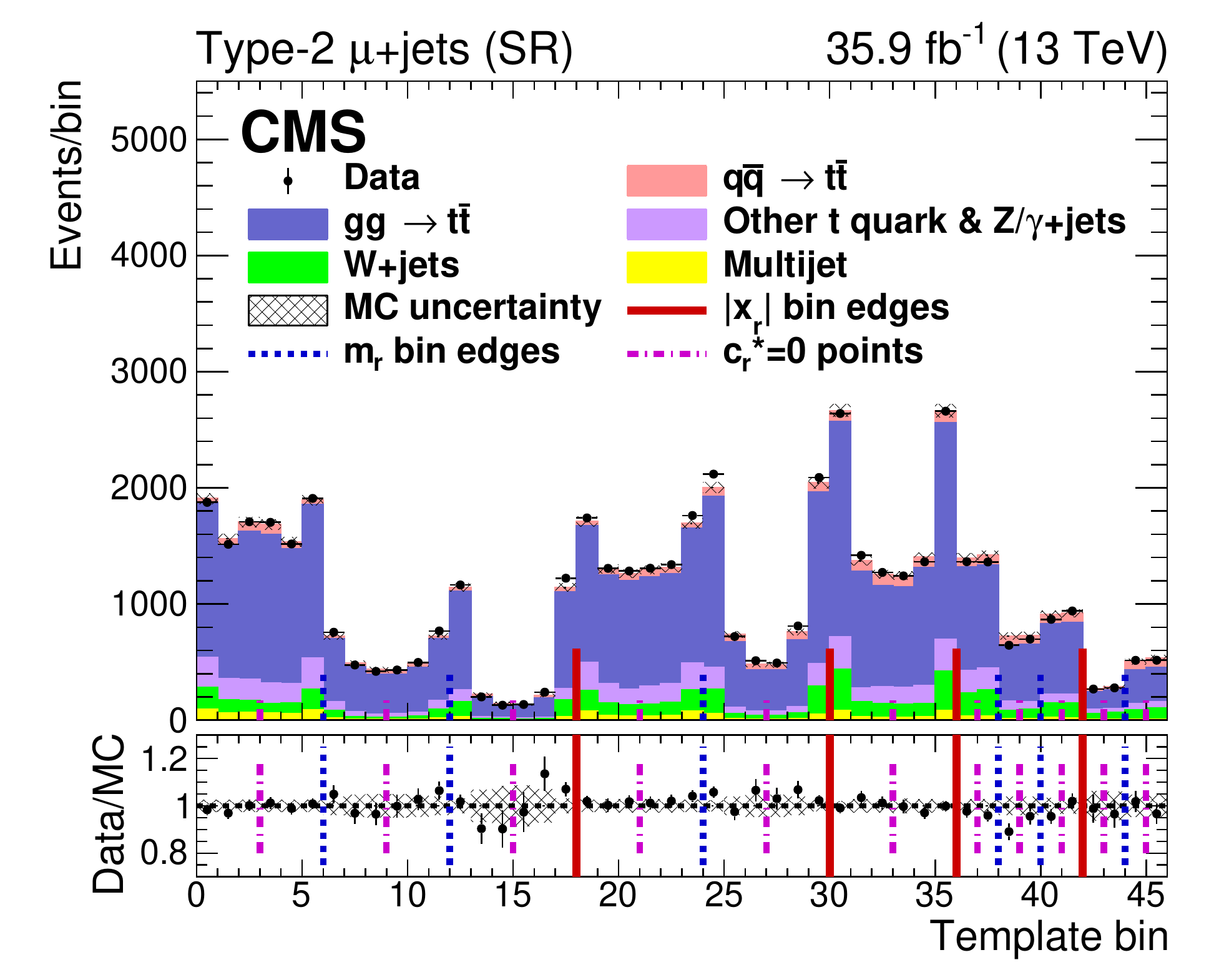}
    \includegraphics[width=0.49\linewidth]{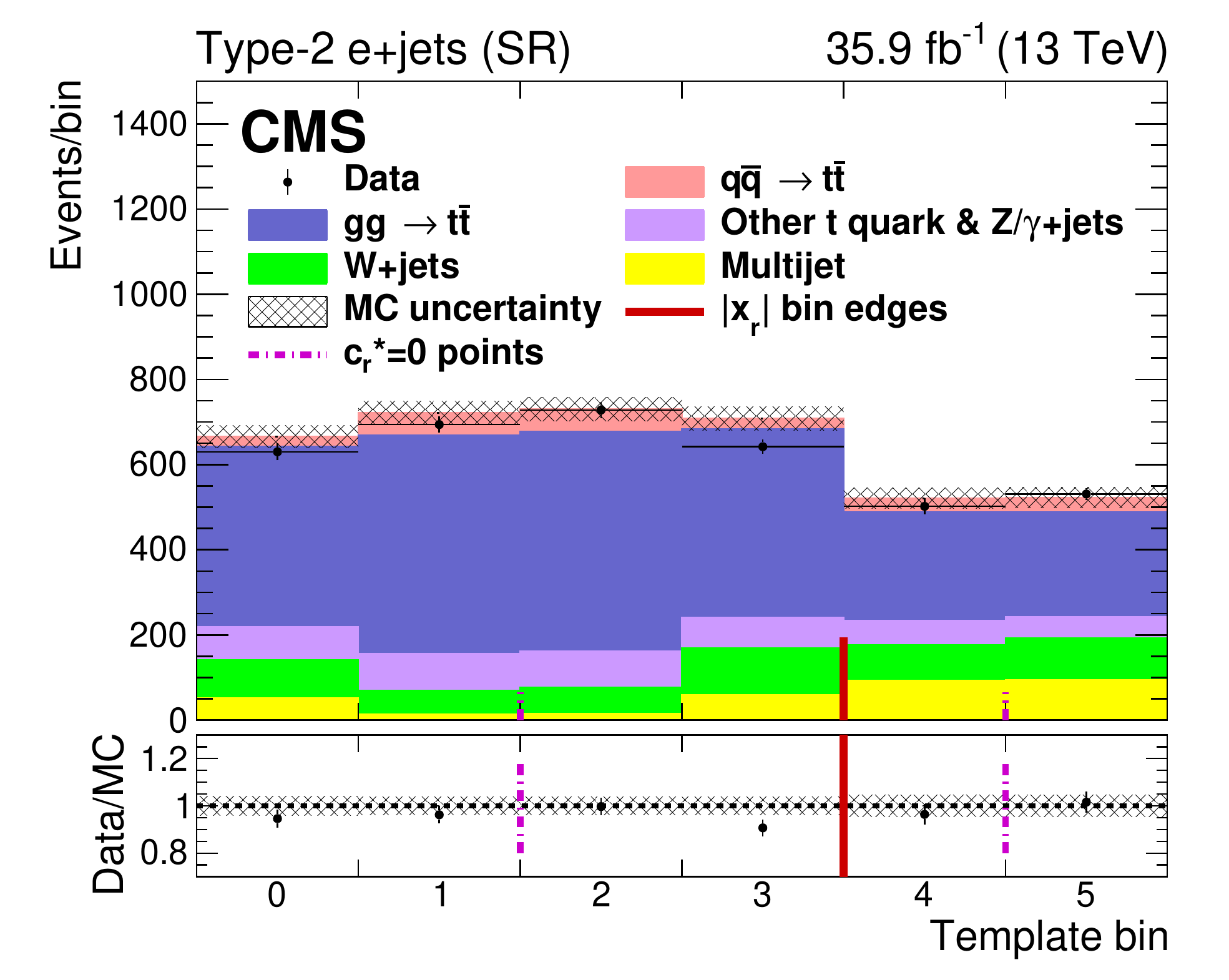}
  \caption{ Comparisons of fitted data and MC expectations as a function of template bin number for the \aofb\ parameter extraction. The plots show events in the type-1 (upper row) and type-2 (lower row) {\PGm}+jets (left column) and {\Pe}+jets (right column) channels, all summed over lepton charge. The MC uncertainty pictured in the hatched bands represents the total (statistical and systematic) uncertainty. The vertical solid, dashed, and dot-dashed lines indicate the edges of the bins in \xr\ and \mr\, and the midpoints of the \cstar\ distributions, respectively, corresponding to the binning schemes listed in Tables~\ref{tab:template_binning_t1_muons}--\ref{tab:template_binning_t2_electrons} in Appendix~\ref{app:template_binning}.  }
    \label{fig:postfit_comp_Afb_boosted_SR}
\end{figure}

\begin{figure}[hbt]
  \centering
    \includegraphics[width=0.98\linewidth]{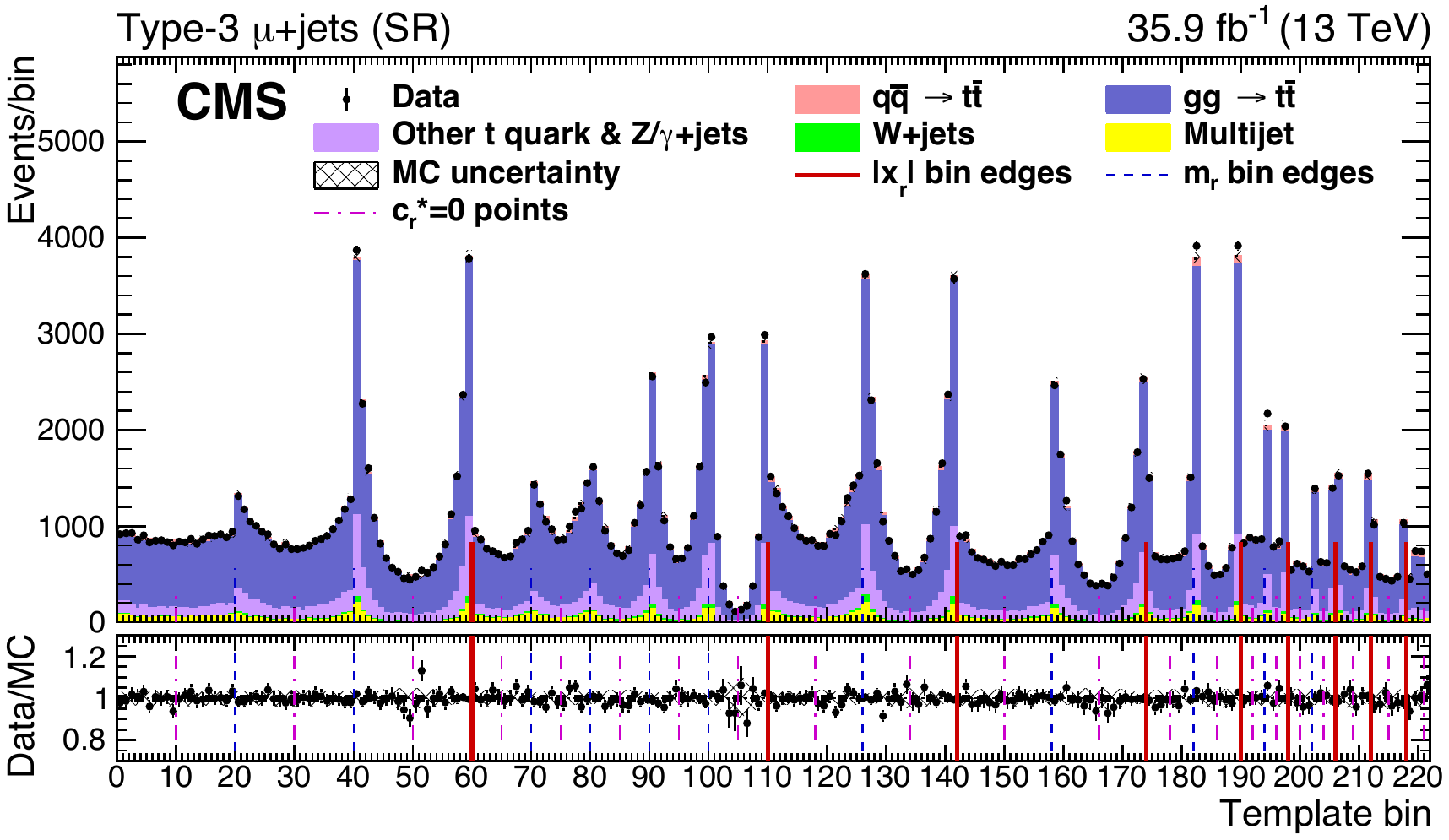}
    \includegraphics[width=0.98\linewidth]{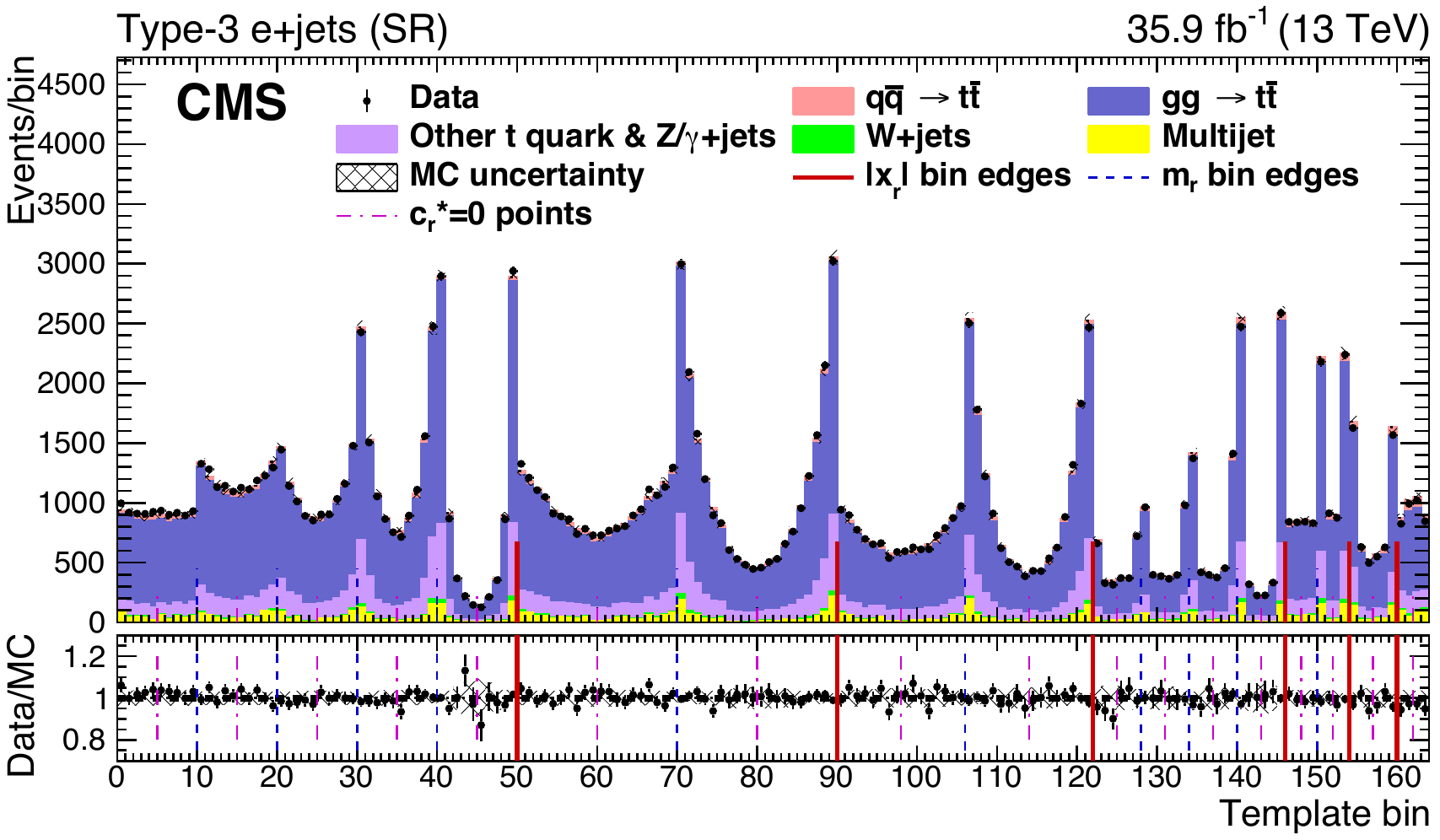}
  \caption{ Comparisons of fitted data and MC expectations as a function of template bin number for the \aofb\ parameter extraction. The plots show events in the type-3 {\PGm}+jets (upper) and {\Pe}+jets (lower) channels, all summed over lepton charge. The MC uncertainty pictured in the hatched bands represents the total (statistical and systematic) uncertainty. The vertical solid, dashed, and dot-dashed lines indicate the edges of the bins in \xr\ and \mr\, and the midpoints of the \cstar\ distributions, respectively, corresponding to the binning schemes listed in Tables~\ref{tab:template_binning_t3_muons} and~\ref{tab:template_binning_t3_electrons} in Appendix~\ref{app:template_binning}.}
    \label{fig:postfit_comp_Afb_resolved_SR}
\end{figure}

The application of the fitting procedure to the data yields $\aofb=0.048^{+0.095}_{-0.087}\stat^{+0.020}_{-0.029}\syst$ and $\parmu=-0.024^{+0.013}_{-0.009}\stat^{+0.016}_{-0.011}\syst$. Figure~\ref{fig:neyman_plots} shows the Neyman construction plots corresponding to these final values. The profile likelihood shown in Fig.~\ref{fig:d_likelihood_profile} results in a 95\% confidence limit $\abs{\pard}<0.03$. Figures~\ref{fig:postfit_comp_Afb_boosted_SR} and~\ref{fig:postfit_comp_Afb_resolved_SR} show comparisons of representative fit functions for the \aofb\ analysis in each signal region (SR) channel, summed over lepton charge. Overall, we observe good agreement, and results from the \pard\ and \parmu\ analyses are comparable.

The goodness of fit is evaluated using a $\chi^2$ test statistic optimized for Poisson-distributed data~\cite{BAKER1984437}.  Comparisons of the statistic observed for each fit to data with those observed in fits to samples of pseudo-data yield $p$-values that quantify the fraction of pseudo-experiments with poorer fits. The $p$-values for the fits extracting \aofb, \pard, and \parmu\ are 0.206, 0.233, and 0.274, respectively, suggesting that the fitting models reasonably describe the observed data.

The measured \aofb\ is consistent with the SM expectation.  The measured anomalous chromoelectric and chromomagnetic moments are consistent with SM expectations ($\parmu = \pard = 0$), and with the most recent results extracted from top quark spin correlation measurements performed by CMS at 13\TeV\ \cite{Sirunyan:2019lnl}.  The spin correlation measurements, which are more precise by a factor of approximately two, use samples of dileptonic final states and are independent of the lepton+jets based measurements presented in this paper.

\clearpage

\section{Summary}
\label{sec:summary}

The linearized parton-level top quark forward-backward asymmetry (\aofb) and anomalous chromoelectric (\pard) and chromomagnetic (\parmu) moments have been measured using LHC proton-proton collisions at a center-of-mass energy of 13\TeV, corresponding to an integrated luminosity of 35.9\fbinv. Candidate \ttbar events decaying to lepton+jets final states with ``resolved" (low-energy) and ``boosted" (high-energy) topologies were selected and reconstructed through a kinematic fit of the decay products to \ttbar hypotheses. The parameters were extracted from independent template-based likelihood fits to the data based on differential models of extensions to leading-order tree-level cross sections for quark-antiquark and gluon-gluon initial states, yielding $\aofb=0.048^{+0.095}_{-0.087}\stat^{+0.020}_{-0.029}\syst$, $\parmu=-0.024^{+0.013}_{-0.009}\stat^{+0.016}_{-0.011}\syst$, and $\abs{\pard}<0.03$ at 95\% confidence level.

\begin{acknowledgments}
We congratulate our colleagues in the CERN accelerator departments for the excellent performance of the LHC and thank the technical and administrative staffs at CERN and at other CMS institutes for their contributions to the success of the CMS effort. In addition, we gratefully acknowledge the computing centers and personnel of the Worldwide LHC Computing Grid for delivering so effectively the computing infrastructure essential to our analyses. Finally, we acknowledge the enduring support for the construction and operation of the LHC and the CMS detector provided by the following funding agencies: BMBWF and FWF (Austria); FNRS and FWO (Belgium); CNPq, CAPES, FAPERJ, FAPERGS, and FAPESP (Brazil); MES (Bulgaria); CERN; CAS, MoST, and NSFC (China); COLCIENCIAS (Colombia); MSES and CSF (Croatia); RPF (Cyprus); SENESCYT (Ecuador); MoER, ERC IUT, PUT and ERDF (Estonia); Academy of Finland, MEC, and HIP (Finland); CEA and CNRS/IN2P3 (France); BMBF, DFG, and HGF (Germany); GSRT (Greece); NKFIA (Hungary); DAE and DST (India); IPM (Iran); SFI (Ireland); INFN (Italy); MSIP and NRF (Republic of Korea); MES (Latvia); LAS (Lithuania); MOE and UM (Malaysia); BUAP, CINVESTAV, CONACYT, LNS, SEP, and UASLP-FAI (Mexico); MOS (Montenegro); MBIE (New Zealand); PAEC (Pakistan); MSHE and NSC (Poland); FCT (Portugal); JINR (Dubna); MON, RosAtom, RAS, RFBR, and NRC KI (Russia); MESTD (Serbia); SEIDI, CPAN, PCTI, and FEDER (Spain); MOSTR (Sri Lanka); Swiss Funding Agencies (Switzerland); MST (Taipei); ThEPCenter, IPST, STAR, and NSTDA (Thailand); TUBITAK and TAEK (Turkey); NASU (Ukraine); STFC (United Kingdom); DOE and NSF (USA).

\hyphenation{Rachada-pisek} Individuals have received support from the Marie-Curie program and the European Research Council and Horizon 2020 Grant, contract Nos.\ 675440, 752730, and 765710 (European Union); the Leventis Foundation; the A.P.\ Sloan Foundation; the Alexander von Humboldt Foundation; the Belgian Federal Science Policy Office; the Fonds pour la Formation \`a la Recherche dans l'Industrie et dans l'Agriculture (FRIA-Belgium); the Agentschap voor Innovatie door Wetenschap en Technologie (IWT-Belgium); the F.R.S.-FNRS and FWO (Belgium) under the ``Excellence of Science -- EOS" -- be.h project n.\ 30820817; the Beijing Municipal Science \& Technology Commission, No. Z181100004218003; the Ministry of Education, Youth and Sports (MEYS) of the Czech Republic; the Deutsche Forschungsgemeinschaft (DFG) under Germany’s Excellence Strategy -- EXC 2121 ``Quantum Universe" -- 390833306; the Lend\"ulet (``Momentum") Program and the J\'anos Bolyai Research Scholarship of the Hungarian Academy of Sciences, the New National Excellence Program \'UNKP, the NKFIA research grants 123842, 123959, 124845, 124850, 125105, 128713, 128786, and 129058 (Hungary); the Council of Science and Industrial Research, India; the HOMING PLUS program of the Foundation for Polish Science, cofinanced from European Union, Regional Development Fund, the Mobility Plus program of the Ministry of Science and Higher Education, the National Science Center (Poland), contracts Harmonia 2014/14/M/ST2/00428, Opus 2014/13/B/ST2/02543, 2014/15/B/ST2/03998, and 2015/19/B/ST2/02861, Sonata-bis 2012/07/E/ST2/01406; the National Priorities Research Program by Qatar National Research Fund; the Ministry of Science and Education, grant no. 14.W03.31.0026 (Russia); the Programa Estatal de Fomento de la Investigaci{\'o}n Cient{\'i}fica y T{\'e}cnica de Excelencia Mar\'{\i}a de Maeztu, grant MDM-2015-0509 and the Programa Severo Ochoa del Principado de Asturias; the Thalis and Aristeia programs cofinanced by EU-ESF and the Greek NSRF; the Rachadapisek Sompot Fund for Postdoctoral Fellowship, Chulalongkorn University and the Chulalongkorn Academic into Its 2nd Century Project Advancement Project (Thailand); the Nvidia Corporation; the Welch Foundation, contract C-1845; and the Weston Havens Foundation (USA). \end{acknowledgments}

\bibliography{auto_generated}

\clearpage
\appendix

\section{Appendix 1: Reweighting factors}
\label{app:reweighting_factors}

The five parameter-independent template functions $f_{\text{gn}}(\vec y)$ defined in Eq.~(\ref{eq:sixparone}) are constructed from fully simulated \glu\ events at NLO using the following weights~\cite{Haberl:1995ek},
\begin{linenomath}\begin{equation}\begin{aligned}
w_{\mathrm{g}0}(m_{\ttbar}^2, \cstar) &= 1 ,\\
w_{\mathrm{g}1}(m_{\ttbar}^2, \cstar) &= \frac{1}{B(\cstar,\beta)\left(1+\varepsilon \beta^2\cstarsq\right)} ,\\
w_{\mathrm{g}2}(m_{\ttbar}^2, \cstar) &= \frac{56}{\left(1-\beta^2\right)A(\cstar,\beta)B(\cstar,\beta)\left(1+\varepsilon \beta^2\cstarsq\right)} ,\\
w_{\mathrm{g}3}(m_{\ttbar}^2, \cstar) &= \frac{4}{\left(1-\beta^2c^2_*\right)A(\cstar,\beta)B(\cstar,\beta)\left(1+\varepsilon \beta^2\cstarsq\right)} ,\\
w_{\mathrm{g}4}(m_{\ttbar}^2, \cstar) &= \frac{8}{A(\cstar,\beta)B(\cstar,\beta)\left(1+\varepsilon \beta^2\cstarsq\right)}\left[\frac{1}{1-\beta^2c^2_*}+\frac{1}{1-\beta^2}+\frac{4(1-\beta^2c^2_*)}{(1-\beta^2)^2}\right] ,
\end{aligned}\end{equation}\end{linenomath}
\noindent where the $\varepsilon$ function is determined from fits to NLO generator level data in 10 bins of $\beta$ and the functions $A$ and $B$ are defined as
\begin{linenomath}\begin{equation}\begin{aligned}
&A(\cstar,\beta) = \frac{7+9\beta^2\cstarsq}{1-\beta^2\cstarsq} ,\\
&B(\cstar,\beta) = \frac{1-\beta^4c^{*4}+2\beta^2(1-\beta^2)(1-\cstarsq)}{2(1-\beta^2\cstarsq)} .
\end{aligned}\end{equation}\end{linenomath}

The three parameter-independent template functions $f_{\text{qn}}(\vec y)$ defined in Eq.~(\ref{eq:sixparone}) are constructed from fully simulated \qqbar\ initiated events at NLO using the following weights,
\begin{linenomath}\begin{equation}\begin{aligned}
w_{\mathrm{q}0}(m_{\ttbar}^2, \cstar) &= 1 , \\
w_{\mathrm{q}1}(m_{\ttbar}^2, \cstar) &= \frac{4}{1+\beta^2\cstarsq+\left(1-\beta^2\right)+\alpha\left(1-\beta^2\cstarsq\right)} , \\
w_{\mathrm{q}2}(m_{\ttbar}^2, \cstar) &= \frac{4}{1+\beta^2\cstarsq+\left(1-\beta^2\right)+\alpha\left(1-\beta^2\cstarsq\right)} \frac{1-\beta^2\cstarsq}{1-\beta^2} .
\end{aligned}\end{equation}\end{linenomath}

The distribution functions are normalized as follows,
\begin{linenomath}\begin{equation}\begin{aligned}
 & \sum_{Q}\int \rd\xr \rd\mr \rd\csubr f_{\text{bk}}(\xr, \mr, \csubr, Q) = 1 , \\
 & \sum_{Q}\int \rd\xr \rd\mr \rd\csubr f_{\mathrm{g}0}(\xr, \mr, \csubr, Q) = 1 , \\
 & \sum_{Q}\int \rd\xr \rd\mr \rd\csubr f_{\mathrm{q}0}(\xr, \mr, \csubr, Q) = 1 ,
 \end{aligned}\end{equation}\end{linenomath}
\noindent and the functions $F_{\text{gn}}$ and $F_{\text{qn}}$ are the integrals of the reweighted functions $f_{\text{gn}}$ and $f_{\text{qn}}$ for $\text{n}>0$,
\begin{linenomath}\begin{equation}\begin{aligned}
 &F_{\text{gn}} = \sum_{Q}\int \rd\xr \rd\mr \rd\csubr f_{\text{gn}}(\xr, \mr, \csubr, Q) ,\\
 &F_{\text{qn}} = \sum_{Q}\int \rd\xr \rd\mr \rd\csubr f_{\text{qn}}(\xr, \mr, \csubr, Q).
\end{aligned} \end{equation}\end{linenomath}
\noindent These are then used to define the normalization factors in Eq.~(\ref{eq:sixparone}),
\begin{linenomath}\begin{equation}\begin{aligned}
 F_{\glu}(\parmu,\pard) =& 1+\parmu(1+\parmu)F_{\mathrm{g}1}+(\parmu^2+\pard^2)(1+\parmu)F_{\mathrm{g}2} \\
 &+(\parmu^2+\pard^2)(1-5\parmu)F_{\mathrm{g}3}+(\parmu^2+\pard^2)^2F_{\mathrm{g}4} , \\
 F_{\qqbar}(\parmu,\pard) =&1+\left(2\parmu+\parmu^2-\pard^2\right)F_{\mathrm{q}1}+\left(\parmu^2+\pard^2\right)F_{\mathrm{q}2} .
\end{aligned}\end{equation}\end{linenomath}

\section{Appendix 2: Template binning}
\label{app:template_binning}

\begin{table}[ht!]
\topcaption{ Template binning in the type-1 {\PGm}+jets signal regions (each template has 12 total bins)}
\centering
\begin{tabular}{ c c c }
\hline
$\abs{\xr}$ bin range & \mr\ bin edges (\GeVns) & \csubr\ bin edges \\
\hline
0.00 $\to$ 0.10 & 350 $\to$ 5550 & $-1.00$, $-0.50$, $-0.25$, 0.00, 0.25, 0.50, 1.00 \\
0.10 $\to$ 0.24 & 350, 1100, 5550 & $-1$, 0, 1 \\
0.24 $\to$ 1.00 & 350  $\to$ 5550 & $-1$, 0, 1 \\
\hline
\end{tabular}
\label{tab:template_binning_t1_muons}
\end{table}

\begin{table}[ht!]
\topcaption{ Template binning in the type-1 {\Pe}+jets signal regions (each template has 8 total bins)}
\centering
\begin{tabular}{ c c c }
\hline
$\abs{\xr}$ bin range & \mr\ bin edges (\GeVns) & \csubr\ bin edges \\
\hline
0.00 $\to$ 0.10 & 350 $\to$ 5550 & 4 even-width bins \\
0.10 $\to$ 0.24 & 350  $\to$ 5550 & $-1$, 0, 1 \\
0.24 $\to$ 1.00 & 350  $\to$ 5550 & $-1$, 0, 1 \\
\hline
\end{tabular}
\label{tab:template_binning_t1_electrons}
\end{table}

\begin{table}[ht!]
\topcaption{ Template binning in the type-2 {\PGm}+jets signal regions (each template has 46 total bins)}
\centering
\begin{tabular}{ c c c }
\hline
$\abs{\xr}$ bin range & \mr\ bin edges (\GeVns) & \csubr\ bin edges \\
\hline
0.00 $\to$ 0.04 & 350, 730, 860, 5650 & $-1.00$, $-0.50$, $-0.25$, 0.00, 0.25, 0.50, 1.00 \\
0.04 $\to$ 0.09 & 350, 730, 5650 & $-1.00$, $-0.50$, $-0.25$, 0.00, 0.25, 0.50, 1.00 \\
0.09 $\to$ 0.15 & 350 $\to$ 5650 & $-1.00$, $-0.50$, $-0.25$, 0.00, 0.25, 0.50, 1.00 \\
0.15 $\to$ 0.24 & 350, 730, 860, 5650 & $-1$, 0, 1 \\
0.24 $\to$ 1.00 & 350, 730, 5650 & $-1$, 0, 1 \\
\hline
\end{tabular}
\label{tab:template_binning_t2_muons}
\end{table}

\begin{table}[ht!]
\topcaption{ Template binning in the type-2 {\Pe}+jets signal regions (each template has 6 total bins)}
\centering
\begin{tabular}{ c c c }
\hline
$\abs{\xr}$ bin range & \mr\ bin edges (\GeVns) & \csubr\ bin edges \\
\hline
0.00 $\to$ 0.15 & 350 $\to$ 5650 & 4 even-width bins \\
0.15 $\to$ 1.00 & 350 $\to$ 5650 & $-1$, 0, 1 \\
\hline
\end{tabular}
\label{tab:template_binning_t2_electrons}
\end{table}

\begin{table}[ht!]
\topcaption{ Template binning in the type-3 {\PGm}+jets signal regions (each template has 222 total bins)}
\centering
\begin{tabular}{ c c c }
\hline
$\abs{\xr}$ bin range & \mr\ bin edges (\GeVns) & \csubr\ bin edges \\
\hline
0.000 $\to$ 0.020 & 350, 425, 500, 2650 & 20 even-width bins \\
0.020 $\to$ 0.040 & 350, 400, 450, 500, 600, 2650 & 10 even-width bins \\
0.040 $\to$ 0.060 & 350, 475, 2650 & 16 even-width bins \\
0.060 $\to$ 0.080 & 350, 475, 2650 & 16 even-width bins \\
0.080 $\to$ 0.100 & 350, 475, 2650 & $-1.0$, $-0.6$, $-0.4$, $-0.2$, 0.0, 0.2, 0.4, 0.6, 1.0 \\
0.100 $\to$ 0.115 & 350, 475, 2650 & 4 even-width bins \\
0.115 $\to$ 0.130 & 350, 475, 2650 & 4 even-width bins \\
0.130 $\to$ 0.150 & 350 $\to$ 2650 & $-1.00$, $-0.50$, $-0.25$, 0.00, 0.25, 0.50, 1.00 \\
0.150 $\to$ 0.180 & 350 $\to$ 2650 & $-1.00$, $-0.50$, $-0.25$, 0.00, 0.25, 0.50, 1.00 \\
0.180 $\to$ 1.000 & 350 $\to$ 2650 & 4 even-width bins \\
\hline
\end{tabular}
\label{tab:template_binning_t3_muons}
\end{table}

\begin{table}[ht!]
\topcaption{ Template binning in the type-3 {\Pe}+jets signal regions (each template has 164 total bins)}
\centering
\begin{tabular}{ c c c }
\hline
$\abs{\xr}$ bin range & \mr\ bin edges (\GeVns) & \csubr\ bin edges \\
\hline
0.000 $\to$ 0.025 & 350, 400, 450, 500, 600, 2650 & 10 even-width bins \\
0.025 $\to$ 0.050 & 350, 475, 2650 & 20 even-width bins \\
0.050 $\to$ 0.075 & 350, 475, 2650 & 16 even-width bins \\
0.075 $\to$ 0.100 & 350, 420, 475, 550, 2650 & $-1.00$, $-0.50$, $-0.25$, 0.00, 0.25, 0.50, 1.00 \\
0.100 $\to$ 0.125 & 350, 475, 2650 & 4 even-width bins \\
0.125 $\to$ 0.155 & 350 $\to$ 2650 & $-1.00$, $-0.50$, $-0.25$, 0.00, 0.25, 0.50, 1.00 \\
0.155 $\to$ 1.000 & 350 $\to$ 2650 & 4 even-width bins \\
\hline
\end{tabular}
\label{tab:template_binning_t3_electrons}
\end{table}

\cleardoublepage \section{The CMS Collaboration \label{app:collab}}\begin{sloppypar}\hyphenpenalty=5000\widowpenalty=500\clubpenalty=5000\vskip\cmsinstskip
\textbf{Yerevan Physics Institute, Yerevan, Armenia}\\*[0pt]
A.M.~Sirunyan$^{\textrm{\dag}}$, A.~Tumasyan
\vskip\cmsinstskip
\textbf{Institut f\"{u}r Hochenergiephysik, Wien, Austria}\\*[0pt]
W.~Adam, F.~Ambrogi, T.~Bergauer, M.~Dragicevic, J.~Er\"{o}, A.~Escalante~Del~Valle, M.~Flechl, R.~Fr\"{u}hwirth\cmsAuthorMark{1}, M.~Jeitler\cmsAuthorMark{1}, N.~Krammer, I.~Kr\"{a}tschmer, D.~Liko, T.~Madlener, I.~Mikulec, N.~Rad, J.~Schieck\cmsAuthorMark{1}, R.~Sch\"{o}fbeck, M.~Spanring, W.~Waltenberger, C.-E.~Wulz\cmsAuthorMark{1}, M.~Zarucki
% ...
\vskip\cmsinstskip
\dag: Deceased\\
1:  Also at Vienna University of Technology, Vienna, Austria\\
% ...
\end{sloppypar}
\end{document}